\documentclass[12pt]{article}

\usepackage{amssymb}
\usepackage{amsmath}
\usepackage{latexsym}
\usepackage{epsfig}
\usepackage{graphicx}
\usepackage{subfigure}
\usepackage{wasysym}
\usepackage{threeparttable}
\usepackage{booktabs}
\usepackage[round]{natbib}
\usepackage{color}
\usepackage{epstopdf}
\usepackage{bm}
\usepackage{tabularx}
\usepackage{verbatim}
\usepackage{hyperref}
\usepackage{float}
\usepackage[ruled,vlined]{algorithm2e}
\usepackage{ulem}
\usepackage{listings}
\usepackage{color}
\usepackage[usenames,dvipsnames]{xcolor}
\usepackage{longtable}

\def\boxit#1{\vbox{\hrule\hbox{\vrule\kern6pt
          \vbox{\kern6pt#1\kern6pt}\kern6pt\vrule}\hrule}}

\def\bse{\begin{eqnarray*}}
\def\ese{\end{eqnarray*}}
\def\be{\begin{eqnarray}}
\def\ee{\end{eqnarray}}
\def\bq{\begin{equation}}
\def\eq{\end{equation}}
\def\bse{\begin{eqnarray*}}
\def\ese{\end{eqnarray*}}

\def\boxit#1{\vbox{\hrule\hbox{\vrule\kern6pt
          \vbox{\kern6pt#1\kern6pt}\kern6pt\vrule}\hrule}}

\def\bfg{\mathbf{g}}

\def\bfu{\mathbf{u}}

\def\bfx{\mathbf{x}}
\def\bfy{\mathbf{y}}

\def\bfI{\mathbf{I}}

\newcommand{\bftheta}{\mbox{\boldmath $\theta$}}

\newcommand{\bfepsilon}{\mbox{\boldmath $\varepsilon$}}
\newcommand{\bfPhi}{\mbox{\boldmath $\Phi$}}
\newcommand{\bfPsi}{\mbox{\boldmath $\Psi$}}

\newcommand{\bfSigma}{\mbox{\boldmath $\Sigma$}}

\newcommand{\Lagr}{\mathcal{L}}

\begin{document}
\thispagestyle{empty}
\baselineskip=28pt

\begin{center}
{\LARGE{\bf Bayesian Circular Lattice Filters for Computationally Efficient Estimation of Multivariate Time-Varying Autoregressive Models}}

\end{center}

\baselineskip=12pt

\vskip 2mm
\begin{center}
Yuelei Sui\footnote{(\baselineskip=10pt to whom correspondence should be addressed) SAS Institute, North Carolina, USA},
Scott H. Holan\footnote{\baselineskip=10pt
Department of Statistics, University of Missouri, Missouri, USA}\footnote{\baselineskip=10ptOffice of the Associate Director for Research and Methodology, U.S. Census Bureau, Washington, D.C., USA},
   and Wen-Hsi Yang\footnote{\baselineskip=10pt School of Agriculture and Food Sciences, The University of Queensland, Queensland, Australia}\,\footnote{\baselineskip=10pt School of Mathematics and Physics, The University of Queensland, Queensland, Australia},
\\
\end{center}
\vskip 4mm

\baselineskip=12pt

\begin{center}
{\bf Abstract}
\end{center}

Nonstationary time series data exist in various scientific disciplines, including environmental science, biology, signal processing, econometrics, among others. Many Bayesian models have been developed to handle nonstationary time series. The time-varying vector autoregressive (TV-VAR) model is a well-established model for multivariate nonstationary time series. Nevertheless, in most cases, the large number of parameters presented by the model results in a high computational burden, ultimately limiting its usage. This paper proposes a computationally efficient multivariate Bayesian Circular Lattice Filter to extend the usage of the TV-VAR model to a broader class of high-dimensional problems. Our fully Bayesian framework allows both the autoregressive (AR) coefficients and innovation covariance to vary over time. Our estimation method is based on the Bayesian lattice filter (BLF), which is extremely computationally efficient and stable in univariate cases. To illustrate the effectiveness of our approach, we conduct a comprehensive comparison with other competing methods through simulation studies and find that, in most cases, our approach performs superior in terms of average squared error between the estimated and true time-varying spectral density. Finally, we demonstrate our methodology through applications to quarterly Gross Domestic Product (GDP) data and Northern California wind data.

\baselineskip=12pt
\par\vfill\noindent
\par\medskip\noindent

\clearpage\pagebreak\newpage \pagenumbering{arabic}
\baselineskip=24pt

\section{Introduction}\label{intro}  
Multivariate time series data are measured and recorded for inquiries of interest in subject-matter disciplines such as biology, ecology, economics, finance, and medicine. For example, multivariate nonstationary time series models work well for modeling correlated economic indicators by using time-varying parameters to evaluate the effects of policy changes and the resulting private sector behavioral changes. In addition, these models are also well-suited for measuring the effect of policy changes on other factors of the economy \citep{huerta2000bayesian,primiceri2005time,nakajima2011bayesian,hunter2017multivariate}. Another example in which these models are useful is multi-channel electroencephalography (EEG) data, where the data are analyzed through their time-frequency representation to reveal how the neuronal activity in one area of the human brain may influence another \citep{ombao2005slex,zhao2019effcient,zhao2022efficient}.
In the era of big data, the number of series and series length within a multivariate time series have been increasing due to technological advances. Consequently, there has also been a fundamental increase in data complexity. Thus, developing efficient methods that scale to such an enormous amount of time series data is imperative.

Parametric and nonparametric approaches for analyzing multivariate time series have been developed to reveal features both between multiple time series and within a single time series. Some parametric methods, e.g., vector autoregressive (VAR) models, are designed for stationary time series. Time-varying vector autoregressive (TV-VAR) models and many other parametric models \citep{ombao2005slex,kowal2017bayesian}, deal with multivariate nonstationary time series. There are also nonparametric methods, e.g., multivariate time-dependent spectral analysis \citep{guo2006multivariate} and 
multivariate polynomial regression \citep{masry1996multivariate,fan2008nonlinear}.

Parametric models for nonstationary multivariate time series play an increasingly important role when modern techniques make high-dimensional data available for analysis. They have several advantages: 1) easy to make forecasts, 2) straightforward to build assessment of uncertainty, and 3) concise descriptions of the underlying model scheme. The TV-VAR model is arguably the most widely used model among the parametric methods. In the time domain, such models have been developed and applied to correlated economic and financial data \citep{primiceri2005time,del2015time,nakajima2011bayesian,nakajima2013bayesian}. In contrast, in the frequency domain, the spectral density and the coherence can often reveal features of each time series that are not readily apparent in the time domain. For example, frequency domain approaches to multi-channel EEG data can aid in the understanding of connective activities within the brain by estimating their spectrum and coherence \citep{ombao2001auto,ombao2005slex,zhao2019effcient,zhao2022efficient}.

Some of the current methods for TV-VAR models assume that the standard deviations of the innovations evolve as geometric random walks and, therefore, belong to the class of stochastic volatility models \citep{shephard2005stochastic, primiceri2005time,del2015time,nakajima2011bayesian}. In the Bayesian setting, these methods typically require Markov chain Monte Carlo (MCMC) methods for estimation and are, therefore, computationally expensive. Alternatively, other methods assume that the innovation variance follows a random walk process. \cite{gersch1995multivariate} proposed an efficient closed-form method to Bayesian inference. Their method used a two-stage estimation approach and assumed a constant innovation covariance. The multivariate dynamic linear models (MDLMs) proposed by \cite{west1997bayesian} allow both the coefficients and the innovation variance to change over time and have been applied to multivariate economic index data \citep{primiceri2005time,del2015time,nakajima2011bayesian}. However, this method is computationally expensive due to sequential computation of matrix inversions.

\cite{zhao2019effcient} extended the Bayesian lattice filter (BLF) of \citet{yang2016bayesian} to TV-VAR models with a constant innovation covariance by implementing the multivariate Durbin--Levison algorithm \citep{brockwell1991time} and using MDLMs on the forward and backward time-varying partial autocorrelation coefficient matrices (TV-VPARCOR). Additionally, \cite{zhao2022efficient} introduced a shrinkage prior and variational approach in the TV-VPARCOR to deal with overfitting. The constant innovation covariance was estimated using an approximation \citep{triantafyllopoulos2007covariance}. However, the assumption of constant innovation covariance may not be reasonable for many types of data. Moreover, these approaches need to calculate inverse matrices sequentially and the computation time still increases exponentially with the increase of the dimension of the data.

To overcome the drawbacks of the high computation cost and the constant innovation covariance assumption, we propose an approach that uses a ``one channel at-a-time" scheme \citep{pagano1978periodic} to transform the multivariate time series model into a periodic univariate process. This ``one channel at-a-time" algorithm efficiently reduces the computation cost and makes parallel computing possible for high-dimensional data. \cite{sakai1982circular} developed a circular lattice structure to estimate the AR coefficients iteratively based on \cite{pagano1978periodic}. This approach allows us to estimate the AR coefficients one series at a time and, therefore, minimizes the need to calculate big matrices. This strategy makes the computational cost increase linearly, rather than exponentially, with the model order. \cite{gersch1995multivariate} applied this circular lattice to the time-varying multivariate AR model using a smoothness prior \citep{kitagawa1996smoothness} on the AR coefficients. This method uses a two-stage estimation approach for the coefficients and innovation covariance and only allows coefficients to be time-varying. The assumption of constant innovation covariance limits application for many types of data. Our new approach uses dynamic linear models (DLMs) to facilitate the estimation in each stage of the lattice structure, allowing both the coefficients and the innovation variances to vary with time  (see also, \cite{sui2021nonstationary}). By modeling the time-varying innovation covariance, the resulting models are more broadly applicable. Additionally, we adopt the methods by \cite{levy2021dynamic} to address the problem of ordering uncertainty within the multivariate time series. 

The remainder of the paper is organized as follows. First, we introduce Bayesian circular lattice filters in Section~\ref{sec:methodology} and evaluate the method via extensive simulation studies in Section~\ref{sec:simulation}. In Section~\ref{sec:application}, we apply the method to two applications, modeling quarterly GDP for five countries and modeling wind speed from three stations in Northern California. Finally, Section~\ref{sec:discussion} contains conclusion and discussion. Comprehensive details and algorithms for parameter estimation, model selection, and forecasting are provided in the Appendix.

\section{Methodology}\label{sec:methodology}
The Bayesian circular lattice filters (BCLFs) we propose are computationally efficient for order identification and parameter estimation of TV-VAR models with time-varying innovation covariances. This approach takes advantages of a ``one channel at-a-time" scheme and Bayesian lattice structure to achieve its computational efficiency. To introduce this approach, we begin with a general description of TV-VAR models.

\subsection{Time-varying Vector Autoregressive Model and Bayesian Inference}
Suppose an observed $K$-variate time series of length $T$ is defined as $\bfx_t = (x_{1,t},\dots, x_{K,t})'$ for $t = 1,\dots, T$. For this time series, we define the time-varying vector AR model with order $P$
(TV-VAR($P$)) as 
\begin{equation}\label{eq:1}
    \bfx_t = \overset{P}{\underset{p=1}{\sum}} \bfPhi_{p,t}\bfx_{t-p} + \bfu_t, \quad \bfu_t \sim N(0, \bfSigma_t),
\end{equation}
where $\bfPhi_{p,t}$ and $\bfu_t$ are the $K \times K$ AR coefficient matrix for lag $p$ and the $K \times 1$ innovation vector at time $t$, respectively. The innovation $\bfu_t$ is assumed to follow a multivariate Gaussian distribution with zero-mean and time-dependent covariance matrix, $\bfSigma_t$. This definition for the TV-VAR model results in nonstationary behavior as a consequence of the coefficient and innovation covariance matrices varying over time.

Among many Bayesian inference approaches to TV-VAR models, the elements of $\bfPhi_{p,t}$ are often assumed to follow random walk processes over time. The prior distributions are often assumed on the decomposed innovation covariance matrix.  After applying an LDL decomposition (also called modified Cholesky decomposition) on the innovation covariance matrix, there are different assumptions on the elements of the diagonal matrix D. The stochastic volatility class of TV-VAR models \citep{primiceri2005time,del2015time,nakajima2011bayesian} assume the logarithm of the diagonal elements of the innovation covariance follow a random walk or an AR process. As such, these models can capture changing innovation variances. Alternatively, MDLMs provide full posterior inference for TV-VAR parameters and assume the diagonal elements of the innovation covariance follow random walk processes. However, the inference of MDLMs requires expensive matrix computations. Thus, the usage is limited to those time series with a small number of series and TV-VAR models of low orders. \cite{zhao2019effcient} proposed a multivariate BLF in which the computational cost increases linearly with the model order, but the computational cost of their approach still increases exponentially with the number of series. In contrast, we propose methods that avoid cumbersome matrix calculations in the BLFs. 

\subsection{Time-varying Periodic Time Series}
\cite{pagano1978periodic} proposed a ``one-channel-at-a-time" modeling approach to break the VAR model into scalar periodic AR processes. This approach makes the computational cost increase linearly rather than exponentially with the model order. For a $K$-variate TV-VAR model such as \eqref{eq:1}, we consider an LDL decomposition such that $\bfSigma_t=L_t W_tL'_t$, where $L_t$ is a lower unit triangular matrix and $W_t$ is a diagonal matrix. With both sides of \eqref{eq:1} premultiplied by the inverse of $L_t$, the covariance matrix reduces to a diagonal matrix $W_t$, such that $W_t = \text{diag}(\sigma^2_{1,t},\dots,\sigma^2_{K,t})$. As such, we obtain the instantaneous response-orthogonal innovations model \citep{gersch1994one,gersch1995multivariate,kitagawa1996smoothness}
\begin{equation}\label{eq:2}
 L^{-1}_t\bfx_t = \sum_{p=1}^{P}A_{p,t} \bfx_{t-p} + \bfepsilon_t,~~~\bfepsilon_t \sim N(0, W_t), 
\end{equation}
where $A_{p,t} = L^{-1}_t \bfPhi_{p,t}$ and $\epsilon_{t} = L^{-1}_{t}u_{t}$. \eqref{eq:2} is the model we actually estimate. Since $W_t$ is diagonal, \eqref{eq:2} can be modeled as a set of uncorrelated AR processes. By interlacing the multiple series of $\bfx_t$, we obtain an equivalent periodic TV-AR model, in which it is defined that $y_{k+(t-1)K}=x_{k,t}$, $k=1,\dots,K$, $t=1,\dots,T$. This $K$-channel periodic TV-AR model is given as
\begin{equation}\label{eq:3}
    y_{k+(t-1)K} = \sum_{m=1}^{M_k} a_{m,k+(t-1)K} y_{k+(t-1)K-m} + \epsilon_{k+(t-1)K},
\end{equation}
where $a_{m,k+(t-1)K}$ is the $m$th AR coefficient at time $t$, $\epsilon_{k+(t-1)K}$ is the innovation with variance $\sigma^2_{k+(t-1)K}=\sigma^2_{k,t-1}$ and $M_k=KP+k-1$ is the order of the $k$th series, $y_{k+(t-1)K}$, for $t=1,\dots,T$. Therefore, the TV-VAR model given by \eqref{eq:1} can be rewritten as a periodic TV-AR model defined by \eqref{eq:3}. The periodic TV-AR can be expressed in a $K$-vector form. For example, for $K$ = 3, $P$ = 2, \eqref{eq:3} can be written in matrix form as
\begin{gather}\label{eq:matrix}
\begin{aligned}
\begin{bmatrix}
            1  &          0   &  0 \\
  -a_{1,3(t-1)+2}  &         1    &  0\\
   -a_{2,3(t-1)+3} &  -a_{1,3(t-1)+3} &  1
 \end{bmatrix} 
 \begin{bmatrix}
  y_{3(t-1)+1} \\
  y_{3(t-1)+2} \\
  y_{3(t-1)+3} 
 \end{bmatrix}  = &
  \begin{bmatrix}
  a_{3,3(t-1)+1}  & a_{2,3(t-1)+1} &  a_{1,3(t-1)+1} \\
  a_{4,3(t-1)+2}  & a_{3,3(t-1)+2} &  a_{2,3(t-1)+2} \\
  a_{5,3(t-1)+3}  & a_{4,3(t-1)+3} &  a_{3,3(t-1)+3} 
 \end{bmatrix}
 \begin{bmatrix}
  y_{3(t-1)-2} \\
  y_{3(t-1)-1} \\
  y_{3(t-1)} 
 \end{bmatrix} \\
  +\begin{bmatrix}
  a_{6,3(t-1)+1}  & a_{5,3(t-1)+1}   & a_{4,3(t-1)+1} \\
  a_{7,3(t-1)+2} & a_{6,3(t-1)+2}  &  a_{5,3(t-1)+2} \\
   a_{8,3(t-1)+3} &  a_{7,3(t-1)+3}  & a_{6,3(t-1)+3} 
 \end{bmatrix}
& \begin{bmatrix}
  y_{3(t-1)-5} \\
  y_{3(t-1)-4} \\
  y_{3(t-1)-3} 
 \end{bmatrix} +
   \begin{bmatrix}
  \epsilon_{3(t-1)+1} \\
  \epsilon_{3(t-1)+2} \\
  \epsilon_{3(t-1)+3} 
 \end{bmatrix}, \quad   t=1,\dots,T.
\end{aligned}
\end{gather}
As in \eqref{eq:matrix}, each channel has its own process. As such, the $K$-channel periodic TV-AR process is named in the sense that it is obtained by interlacing $K$ uncorrelated TV-AR processes.

The parameters in \eqref{eq:1} and the parameters in the corresponding periodic TV-AR model given by \eqref{eq:3} have the relationship
 \begin{align}
     \bfPhi_{p,t} &= -L_t A_{p,t} \label{eq:A1}\\
     \Sigma_t &= L_t W_t L'_t.   \label{eq:A2} 
 \end{align}
This relationship allows us to rewrite a TV-VAR model as a periodic TV-AR model, and vice versa.

\subsection{Circular Lattice Filter}
We propose a BCLF model, which can handle both time-varying coefficients and time-varying innovation covariance. According to the Durbin-Levinson algorithm, there exists a unique correspondence between the PARCOR coefficients and the AR coefficients,  \citep{shumway2006time,kitagawa2010timeseries,yang2016bayesian}. This provides a direct way of estimating AR models through the PARCOR coefficients (see \cite{hayes1996sdspam} and the Supplementary Appendix of \cite{yang2016bayesian}). Here, we modify the above approach for TV-VAR models. In the periodic TV-AR model defined by \eqref{eq:3}, the series is fitted by TV-AR models iteratively. The PARCOR coefficients are estimated through TV-AR models in every stage of the circular lattice structure. Finally, the PARCOR coefficients are transformed into TV-VAR coefficients. We denote $f^{(M_k)}_{k+(t-1)K}$ and $b^{(M_k)}_{k+(t-1)K}$ to be the prediction error of $k$th series at time $t$ for the forward and backward time-varying AR$(M_k)$ models, respectively, where
  \begin{align*}
    f^{(M_k)}_{k+(t-1)K} &= y_{k+(t-1)K} - \sum_{m=1}^{M_k} a_{m,k+(t-1)K}^{(M_k)} y_{k+(t-1)K-m}\\
    b^{(M_k)}_{k+(t-1)K} &= y_{k+(t-1)K} - \sum_{m=1}^{M_k} d_{m,k+(t-1)K}^{(M_k)} y_{k+(t-1)K+m}
 \end{align*}
 and $a_{m,k+(t-1)K}^{(M_k)}$ and $b_{m,k+(t-1)K}^{(M_k)}$ are the forward and backward AR coefficients of the corresponding time-varying periodic AR$(M_k)$ models. Then, at the $m$th stage of the lattice filter, for $m=1,\dots,M_k$, the forward and the backward coefficients and the forward and  backward prediction errors have the relationship
   \begin{align}
     f_{k+(t-1)K}^{(m)} &= f_{k+(t-1)K}^{(m-1)} - \alpha_{f,m,k+(t-1)K}^{(m)} b_{k-1+tK}^{(m-1)}, \label{eq:fb1}\\
     b_{k+(t-1)K}^{(m)} &= b_{k-1+tK}^{(m-1)} - \alpha_{b,m,k+(t-1)K}^{(m)} f_{k+(t-1)K}^{(m-1)}, \label{eq:fb2}
 \end{align}
with the initial condition, $f^{(0)}_{k+(t-1)K} = b^{(0)}_{k+(t-1)K} = y_{k+(t-1)K}$, and where $\alpha_{f,m,k+(t-1)K}^{(m)}$ and $\alpha_{b,m,k+(t-1)K}^{(m)}$ are the lag $m$ forward and backward PARCOR coefficients of the $k$th series at time $t$, respectively. The $k$th series $j$th lag forward and backward AR coefficients at time $t$, $a_{j,k+(t-1)K}^{(m)}$ and $d_{j,k+(t-1)K}^{(m)}$, can be obtained according to the following equations
\begin{equation*}
  \begin{aligned}
     a_{j,k+(t-1)K}^{(m)} &= a_{j,k+(t-1)K}^{(m-1)} - a_{m,k+(t-1)K}^{(m)} d_{m-j,k+(t-1)K}^{(m-1)},\\
     d_{j,k+(t-1)K}^{(m)} &= d_{j,k+(t-1)K}^{(m-1)} - d_{m,k+(t-1)K}^{(m)} a_{m-j,k+(t-1)K}^{(m-1)},
 \end{aligned}
\end{equation*}
 with $j = 1,\dots, m-1$, $a_{m,k+(t-1)K}^{(m)} = \alpha_{f,m,k+(t-1)K}^{(m)}$ and $d_{m,k+(t-1)K}^{(m)} = \alpha_{b,m,k+(t-1)K}^{(m)}$. Going from the first stage through the end, $\{a_{m,k+(t-1)K}^{(M_k)}\}$, $t = 1,\dots,T$, $m=1,\dots,M_k$, are the coefficients in \eqref{eq:3}. The estimated AR coefficients are then obtained through \eqref{eq:A1} and \eqref{eq:A2}. For the $k$th series, when the true process is TV-AR$(M_k)$, the forward innovation variance at stage $M_k$, $\{(\sigma^{(M_k)}_{f,k+(t-1)K})^2\}$ is equal to the innovation variance $\{\sigma^2_{k,t}\}$ for $t = 1,\dots,T$.

\subsection{Model Specification and Bayesian Inference}
To estimate the forward and backward PARCOR coefficients and innovation variances in \eqref{eq:fb1} and \eqref{eq:fb2}, we assume random walks for the PARCOR coefficients and multiplicative random walks for the innovation variance. The PARCOR coefficients are modeled as
 \begin{align*}
     \alpha_{f,m,k+tK}^{(m)} = \alpha_{f,m,k+(t-1)K}^{(m)} + \epsilon_{f,m,k+tK}, ~~~\epsilon_{f,m,k+tK} \sim N(0,\zeta_{f,m,k+tK}),\\
     \alpha_{b,m,k+tK}^{(m)} = \alpha_{b,m,k+(t-1)K}^{(m)} + \epsilon_{b,m,k+tK}, ~~~\epsilon_{b,m,k+tK} \sim N(0,\zeta_{b,m,k+tK}),
 \end{align*}
where $\zeta_{f,m,k+tK}$ and $\zeta_{b,m,k+tK}$ are time dependent evolution variances. These evolution variances are defined via the \textit{discount factors} $\gamma_{f,k,m}$ and $\gamma_{b,k,m}$ within the range $(0,1)$, respectively (see \cite{west1997bayesian} and the Supplementary Appendix of \cite{yang2016bayesian} for details). The discount factor $\gamma$ controls the smoothness of PARCOR coefficients. Here, we assume $\gamma_{f,k,m}= \gamma_{b,k,m} = \gamma_{k,m}$ at each stage $m$ and select their value through a grid-search based on the likelihood of the fitted TV-AR model at each stage. Similarly, the forward and backward innovation variances are modeled as
\begin{align*}
    \sigma_{f,m,k+tK}^2 = \sigma_{f,m,k+(t-1)K}^2(\delta_{f,m}/\eta_{f,m,k+tK}), ~~~ \eta_{f,m,k+tK}\sim Beta(g_{f,m,k+tK},h_{f,m,k+tK}),\\
    \sigma_{b,m,k+tK}^2 = \sigma_{b,m,k+(t-1)K}^2(\delta_{b,m}/\eta_{b,m,k+tK}), ~~~ \eta_{b,m,k+tK}\sim Beta(g_{b,m,k+tK},h_{f,m,k+tK}),
\end{align*}
where $\delta_{f,m}$ and $\delta_{b,m}$ are also discount factors in the range (0,1), and the multiplicative innovations, $\eta_{f,m,k+tK}$ and $\eta_{b,m,k+tK}$, follow beta distributions with parameters $(g_{f,m,k+tK}, h_{f,m,k+tK})$ and $(g_{b,m,k+tK}, h_{b,m,k+tK})$, respectively (see \cite{west1997bayesian} and the Supplementary Appendix of \cite{yang2016bayesian} for details). The smoothness of the innovation variance is controlled by both $\gamma$ and $\delta$. Similar to the PARCOR coefficients, we assume $\delta_{f,k,m}=\delta_{b,k,m}=\delta_{k,m}$ at each stage. Note that $\epsilon_{b,m,k+tK}$, $\epsilon_{b,m,k+tK}$, $\eta_{f,m,k+tK}$ and $\eta_{b,m,k+tK}$ are mutually independent and are also independent of any other variables in the model.

We use conjugate normal priors for the forward and backward PARCOR coefficients, so that
\begin{align*}
    p(\alpha_{f,m,k-K}|D_{f,m,k-K}) &\sim N(\mu_{f,m,k-K},C_{f,m,k-K}),\\
    p(\alpha_{b,m,k-K}|D_{b,m,k-K}) &\sim N(\mu_{b,m,k-K},C_{b,m,k-K}),
\end{align*}
where $m=1,\dots,M_k$, $k=1,\dots,K$, $D_{f,m,k-K}$ and $D_{b,m,k-K}$ denotes the information available at the initial time t = 0, $\mu_{f,m,k-K}$ and $C_{f,m,k-K}$ are the mean and the variance of the normal prior distribution. We also specify conjugate initial priors for the forward and backward innovation variance, so that
\begin{align*}
    p(\sigma_{f,m,k-K}^2|D_{f,m,k-K}) &\sim G(\nu_{f,m,k-K}/2,\kappa_{f,m,k-K}/2),\\
    p(\sigma_{b,m,k-K}^2|D_{b,m,k-K}) &\sim G(\nu_{b,m,k-K}/2,\kappa_{b,m,k-K}/2),
\end{align*}
where $G(\cdot, \cdot)$ is the gamma distribution, and $\nu_{f,m,k-K}/2$ and $\kappa_{f,m,k-K}/2$ are the shape and rate parameters of the gamma prior distribution. Typically, we specify these starting values as constants over all stages. In order to reduce the effect of the prior distribution, we choose $\mu_{f,m,k-K}/2$ and $C_{f,m,k-K}$ to be zero and one, respectively and  fix $\nu_{f,m,0}=1$ and set $\kappa_{f,m,k-K}$ to equal to the sample variance of the initial part of each series. Following these settings, we can obtain the DLM sequential filtering and smoothing algorithms \citep{west1997bayesian} to derive the marginal posterior distributions of the forward and backward parameter parameters in \eqref{eq:fb1} and \eqref{eq:fb2}. The detailed algorithm of the sequential filtering and smoothing are given in \ref{sec:appendixb}. 

\subsection{Model Selection}\label{ms}
To apply the BCLF, we need to find not only the optimal order but also the optimal discount factors. The selection of an optimal model order can be done numerically using model selection criterion. We start with setting up a maximal order, $P_{max}$, and start from the first stage of the lattice filter through the $M_{k,max}$-th stage ($M_{k,max}=KP_{max}+k-1$) for each series. At the $m$th stage, for the $k$th series, we search a group of pre-specified sets of discount factors $\{\gamma_{k,m}, \delta_{k,m}\}$ to find the set which maximizes the likelihood of the fitted DLM, and use the corresponding estimated parameters as the result. Having all the optimal discount factors and the corresponding estimated parameters, the model selection criterion for the order $P$ can be computed based on the estimation obtained from the first through the $(PK+K-1)$th stage. According to the model selection criterion values, the optimal order is selected. For model order selection, we consider several criteria: the Bayesian Information Criterion (BIC), the deviance information criterion (DIC) \citep{gelman2013bayesian}, the widely applicable Akaike information criterion (WAIC) \citep{watanabe2010asymptotic} (see \ref{sec:appendixc} for details). In addition to these criteria, the scree plot method proposed in \cite{yang2016bayesian} also provides a good visual tool to assist the model selection in BCLF.

\subsection{Forecasting}\label{sec:forecast}
Having estimated all parameters, we consider $1$-step ahead forecasts of the TV-VAR($P$) model. The $1$-step ahead predictive distribution of the PARCOR coefficients can be obtained according to \cite{west1997bayesian}. We generate $J$ samples for each of the PARCOR coefficients from their one-step-ahead predictive distribution as $\{\alpha_{f,m,k+TK}^{(m)(j)},\alpha_{b,m,k+TK}^{(m)(j)}$ for all $k,m\}$, $j=1,\dots,J$. The samples of the $1$-step ahead prediction of TV-VAR parameters,  $\Phi_{p,T+1}$ can be obtained as $\{ \bfPhi_{p,T+1}^{(j)},j=1,\dots,J\}$, for $p=1,\dots,P$, by transforming the samples of the PARCOR coefficients according to \eqref{eq:A1} and \eqref{eq:A2}. Finally, the $1$-step ahead forecast is can be obtained using 
\begin{equation}\label{eq:ch3f1}
     \bfx_{T+1}^{(j)} =  \overset{P}{\underset{p=1}{\sum}}   \bfPhi_{p,T+1}^{(j)} \bfx_{T+1-p}. 
\end{equation}
We use the posterior mean of $\bfx_{T+1}^{(j)}$ $(j=1,\ldots, J)$ obtained through the samples in \eqref{eq:ch3f1} as the 1-step ahead forecast. This forecast can be easily extended to $h$-steps. The details for forecasting up to $h$-steps ahead can be found in \ref{sec:appendixd}.

\section{Simulation Studies}\label{sec:simulation}
To assess the effectiveness of our method, we conducted similar simulation studies to \cite{zhao2019effcient} and compared our results with results obtained using their approach and a standard TV-VAR($P$) model implementation. The comparison was conducted using spectral analysis and forecasting. In the spectral analysis, results are assessed using the average squared error (ASE) \citep{ombao2001auto} between the true spectral density and the estimated spectral density. In contrast, the mean squared prediction error (MSPE) is used to evaluate the forecasting performance.

\subsection{Simulation 1: Bivariate TV-VAR(2) Process}\label{sec:sim1}
We consider 500 bivariate time series of length $T=1034$ simulated from a TV-VAR model \citep{zhao2019effcient} as follows:
$\bfx_t=\bfPhi_{1,t} \bfx_{t-1} +\bfPhi_{2,t} \bfx_{t-2} +\bfu_t, ~~~\bfu_t \sim \mathcal{N}({\bf 0}, \bfSigma_t) $
with
$$\bfPhi_{1,t}=\begin{pmatrix}
  r_{1,t} \text{cos}(\frac{2\pi}{\lambda_{1,t}}) & \phi_{1,1,2,t}\\
  0  & r_{2,t} \text{cos}(\frac{2\pi}{\lambda_{2,t}})
 \end{pmatrix} ~~ \text{and}~~
  \bfPhi_{2,t}=\begin{pmatrix}
  -r_{1,t}^2  & \phi_{2,1,2,t}\\
  0  & -r_{2,t}^2
 \end{pmatrix}, $$
 where $r_{1,t}=\frac{0.1}{T}t+0.85$, $r_{2,t}= - \frac{0.1}{T}t + 0.95$, $r_{3,t}= \frac{0.2}{T}t - 0.9$, $r_{4,t}= \frac{0.2}{T}t + 0.7$, $\lambda_{1,t}=\frac{15}{T}t+5$, and $\lambda_{2,t}=-\frac{10}{T}t+15$. We consider three different cases for the values of $\phi_{1,1,2,t}$ and $\phi_{2,1,2,t}$ , namely (1) $\phi_{1,1,2}=0$, $\phi_{2,1,2}=0$; (2) $\phi_{1,1,2}=-0.8$, $\phi_{2,1,2}=0$; (3) $\phi_{1,1,2}=r_{3,t}$, $\phi_{2,1,2}=r_{4,t}$. These three cases have covariances $\bfSigma_t = I_2, 2I_2,$ and $3I_2$, where $I_2$ denotes a $2 \times 2$ identity matrix. We also consider cases (4), (5), (6) with the same $\bfPhi_t$ as (1), (2), (3), respectively, but with different covariance $\bfSigma_t$, such that $\Sigma_{1,1,t}=\Sigma_{2,2,t}=1+\frac{t}{T}$, $\Sigma_{1,2,t}=\Sigma_{2,1,t}=0$ for $t=1,\dots,T$.
 The bivariate spectral matrix of this process can be obtained by
 $$  \bfg(t,\omega) = \bfPsi(t,\omega)^{-1} \times \bfSigma_t  \times \bfPsi^\ast(t,\omega)^{-1},$$
 where $\bfPsi(t,\omega)=\bfI_2 - \overset{P}{\underset{p=1}{\sum}}\bfPhi_{p,t}\text{exp}\{-2p\pi i \omega\}$, $\bfSigma_t$ is time-varying innovation covariance and $\ast$ stands for the Hermitian matrix (conjugate transpose matrix). The spectral matrix $\bfg(t,\omega)$ is symmetric and consists of series $g_{11}(t,\omega)$, $g_{22}(t,\omega)$ and $g_{12}(t,\omega)$, representing the spectrum of the first series, the spectrum of the second series and the cross-spectrum between the first and the second series, respectively. The squared coherence between the first and the second series is defined as
  $$  \rho_{12}^2(t,\omega) = \frac{|g_{12}(t,\omega)|^2}{g_{11}(t,\omega) g_{22}(t,\omega) }.$$
 The spectral matrix can be estimated by
 $$  \widehat{\bfg}(t,\omega) = \widehat{\bfPsi}(t,\omega)^{-1} \times \widehat{\bfSigma}_t  \times \widehat{\bfPsi}^\ast(t,\omega)^{-1}, $$where $\widehat{\bfPsi}(t,\omega) = |\bfI_2 - \overset{P}{\underset{p=1}{\sum}}\widehat{\bfPhi}_{p,t}\text{exp}\{-2p\pi i \omega\}|$ and $\widehat{\bfSigma}_t$ are estimated values. The estimated squared coherence $\widehat{\rho}_{12}^2(t,w)$ can be obtained accordingly.
 
 To evaluate the performance in estimating the time-frequency representations, we compare different models by the mean and standard deviations of the ASEs. We calculate ASE for each realization as follows \citep{ombao2001auto}:
\begin{equation*}
     \text{ASE}_n = (TL)^{-1} \sum_{t=1}^{T}\sum_{l=1}^{L} \left( \text{log}(\widehat{\bfg}(t,\omega_l))- \text{log} (\bfg(t,\omega_l)) \right)^2,    
\end{equation*}
 where $n=1,\dots,500$, $w_l$ is the $l$th frequency, and $L$ is the number of frequencies in the time-frequency representation. We denote the average over all realizations as $\overline{ASE}=1/500 \underset{n=1}{\overset{500}{\sum}}ASE_n$.
 
\begin{figure}
\caption{Boxplots of ASEs by three methods in cases 1, 2, 3 when $\bfSigma_t=I_2$. In each plot, the index 1, 2, and 3 denotes TV-VAR, TV-, and BCLF, respectively.}
\label{fig:321}
\begin{center}
\includegraphics[height=6in,width=6in]{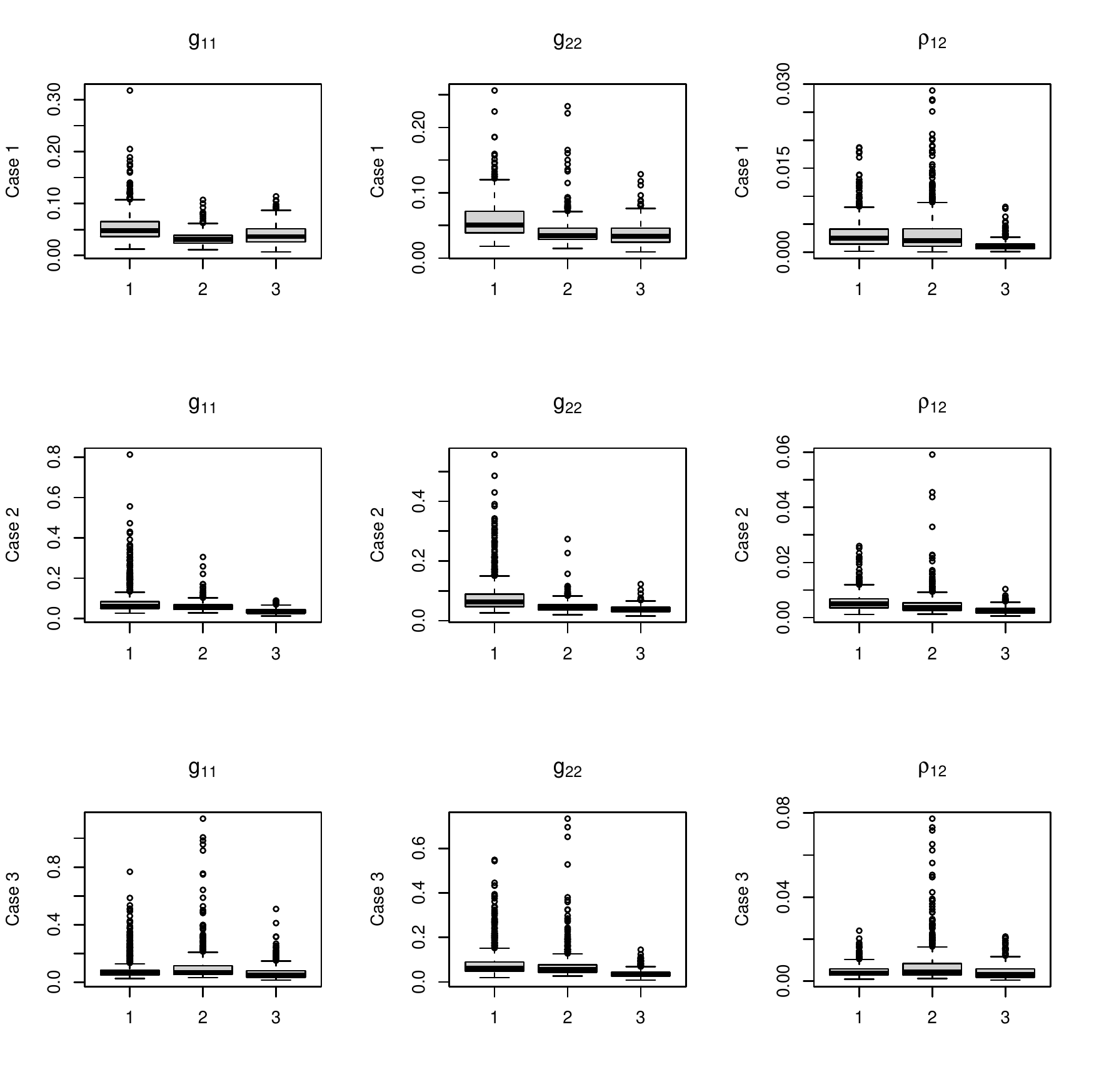}
\end{center}
\end{figure}

\begin{figure}
\caption{Boxplots of ASEs by three methods in cases 4, 5, 6. In each plot, the index 1, 2, and 3 denotes TV-VAR, TV-PARCOR, and BCLF, respectively.}
\label{fig:322}
\begin{center}
\includegraphics[height=6in,width=6in]{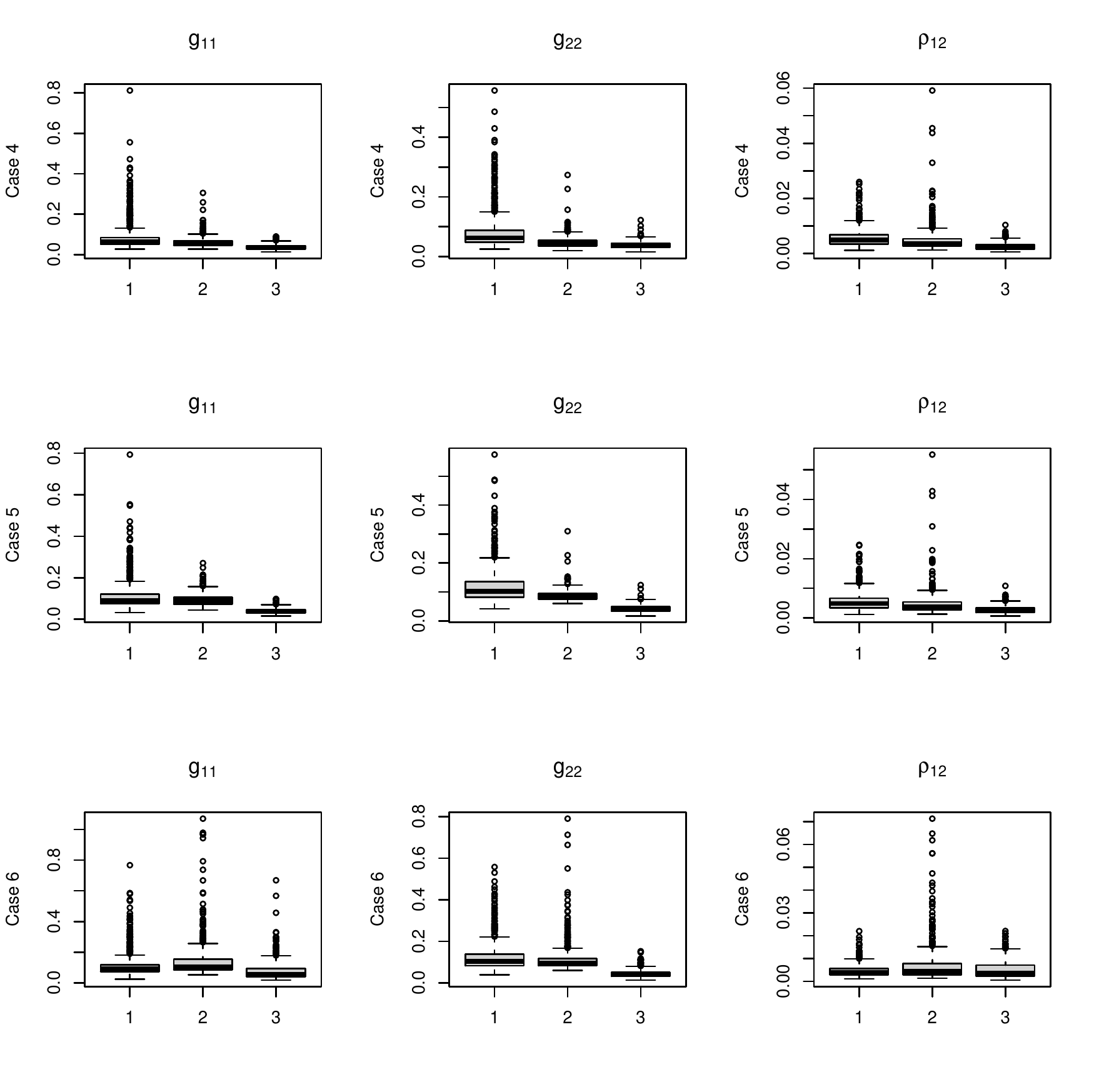}
\end{center}
\end{figure}

\begin{figure}
\caption{Case with $\phi_{1,1,2,t} = -0.8$ and $\phi_{2,1,2,t} = 0$. Top: True log spectral density $g_{11}(t,\omega)$ (left), true log spectral density $g_{22}(t,\omega)$ (middle), true squared coherence $\rho_{12}^2(t,\omega)$ (right). Bottom: Average of the estimated $\widehat{g}_{11}(t,\omega)$ (left), average of the estimated $\widehat{g}_{22}(t,\omega)$ (middle), average of the estimated $\widehat{\rho}_{12}^2(t,\omega)$ (right) using BCLF.}
\label{fig:325}
\begin{center}
\includegraphics[height=6in,width=6in]{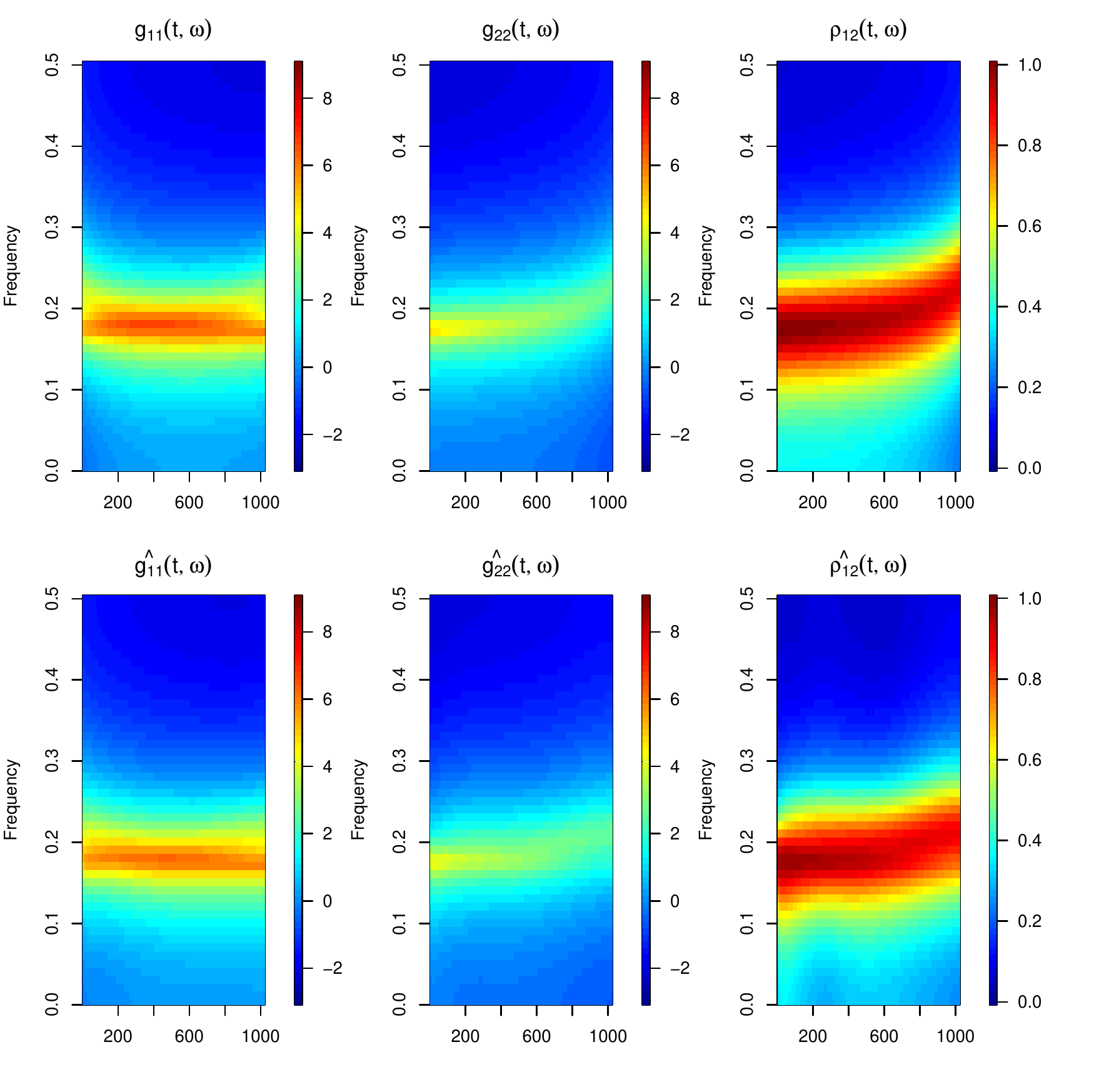}
\end{center}
\end{figure}

To evaluate the performance of our proposed method (BCLF), we choose a $P_{max}=5$ and fit the simulated datasets with the TV-VAR order adaptively selected based on BIC. The other model selection criteria were also obtained and are shown in Table~\ref{table:31}. BIC suggests that TV-VAR(2) is the best model for most simulated datasets. In practice, BIC works extremely well for model selection of the BCLF and significantly better than DIC and WAIC. Consequently, we use BIC for model order selection of the BCLF throughout this paper.
Our method is compared with the TV-VPARCOR and the TV-VAR, as presented in \cite{zhao2019effcient}. As recommended in \cite{zhao2019effcient}, for every dataset, TV-VPARCOR and TV-VAR have the model orders selected based on DIC. In contrast,  BCLF has the model order selected based on BIC. All model selections have a maximum order of 5. The discount factors in all of BCLF, TV-VPARCOR, and TV-VAR are chosen from a grid of values in $[0.99, 1]$ based on the likelihood. We select this range to make our results comparable to those of \cite{zhao2019effcient}. Note that TV-VAR and BCLF consider time-varying covariance while TV-PARCOR assumes constant covariance. Figures~\ref{fig:321} and \ref{fig:322} display boxplots which summarize the ASE for six cases compared with TV-VAR and TV-VPARCOR. Figure~\ref{fig:325} shows the average of the estimated spectrum of the two series and the average of the estimated coherence between them over the 500 simulated datasets in Case (2). 

Further, we compare the forecast of BCLF with TV-PARCOR. The forecast method is given in Sections~\ref{sec:forecast} and \ref{sec:appendixd}. We conduct one-step ahead rolling prediction for $t$ = 1025:1034 for the first 3 simulation cases, which have constant innovation covariances. For each case and each method, 100 series of length 1034 are generated, the last 10 observations of the series are held out for prediction. Table~\ref{table:32} shows the mean and the standard deviation of the 100 MSPEs for each case and each method. In this simulation study, BCLF one-step ahead prediction performs better than TV-VPARCOR.

Table~\ref{table:33} shows the computation time of the simulation studies. The time of Simulation 1 is the time used for fitting one dataset of Case 1, including the selection of discount factors, the selection of the model order, and the estimation of parameters. Note that the TV-VPARCOR method is conducted using Rcpp and parallel computing while the BCLF is coded in R. If TV-VPARCOR is estimated using R (without Rcpp), it would result in  significantly longer computation time than that shown in this table. Moreover, the computation associated with BCLF can be accelerated by using Rcpp and parallel computing, resulting is substantially reduced computation time. In Simulation 1, all of the six cases take approximately the same time and therefore, we use Case 1 as an illustration to represent all the cases.

Each of the methods TV-VAR, TV-VPARCOR, and BCLF have some drawbacks that lower their performance. TV-VAR and TV-VPARCOR use an approximate estimator to estimate the covariance, which does not work as well as our time-varying covariance estimator in some cases. Additionally, TV-VAR uses one discount factor for all coefficients. This produces larger ASEs for the TV-VAR in most cases. The BCLF takes advantage of the modified Cholesky decomposition due to the ``one-channel-at-a-time" algorithm. This decomposition brings the problem of ordering uncertainty \citep{primiceri2005time, zhao2016dynamic, lopes2021parsimony}. Consequently, the priors on the parameters change with the ordering of multiple series and, therefore, the estimated results also change. See Section~\ref{sec:discussion} for additional discussion.

Table~\ref{table:34} shows the ASEs of each method for different coefficients and innovation covariances when the innovation covariances are time-invariant. This table suggests that the BCLF performs better for spectrum estimation in most cases. In other words, the BCLF is more robust to varying levels of the covariance. That is, when the true covariance increases from $1$ to $3$, the ASEs of BCLF stay at approximately the same level, while the ASEs of TV-VAR and TV-VPARCOR become significantly larger. Moreover, Table~\ref{table:35} shows that when the innovation covariance is time-varying, the BCLF works better than the constant covariance models TV-VAR and TV-VPARCOR, as expected. 

\begin{table}
\caption{Model selection of TV-VAR model using BCLF for 500 simulated datasets from bivariate TV-VAR(2) process. Each column gives the model order $P$ and the percentage of the datasets that are selected to this order according to BIC, DIC, and WAIC.}\label{table:31}
\begin{center}
 \begin{tabular}{c | c |c c c c c} 
 \hline
  Case  & Criterion    &  &   & $P$ &  & \\ 

    &     & 1 & 2  & 3 & 4 & 5\\  
 \hline
  & BIC  & 0\% & 100\% & 0\% & 0\% & 0\% \\
1 & DIC  & 0\% & 20\% & 10\% & 23\% & 47\% \\
  & WAIC & 90\% & 10\% & 0\% & 0\% & 0\% \\
 \hline
  & BIC  & 0\% & 100\% & 0\% & 0\% & 1\% \\
2 & DIC  & 0\% & 92\% & 1\% & 4\% & 3\% \\
  & WAIC & 99\% & 1\% & 0\% & 0\% & 0\% \\
 \hline
  & BIC & 0\% & 100\% & 0\% & 0\% & 0\% \\
3 & DIC & 0\% & 87\% & 8\% & 0\% & 5\% \\
  & WAIC & 100\% & 0\% & 0\% & 0\% & 0\% \\
 \hline
  & BIC & 0\% & 100\% & 0\% & 0\% & 0\% \\
4 & DIC & 0\% & 18\% & 11\% & 19\% & 52\% \\
  & WAIC & 100\% & 0\% & 0\% & 0\% & 0\% \\
 \hline
  & BIC & 0\% & 100\% & 0\% & 0\% & 0\% \\
5 & DIC & 0\% & 88\% & 4\% & 0\% & 8\% \\
  & WAIC & 100\% & 0\% & 0\% & 0\% & 0\% \\
 \hline
  & BIC & 0\% & 100\% & 0\% & 0\% & 0\% \\
6 & DIC & 0\% & 82\% & 10\% & 2\% & 6\% \\
  & WAIC & 100\% & 0\% & 0\% & 0\% & 0\% \\
 \hline
\end{tabular}
\end{center}
\end{table}

\begin{table}
\caption{MSPE values for the 1-step ahead rolling forecast ($t$ = 1025:1034) and corresponding standard deviations (in parentheses) obtained from BCLF and TV-VPARCOR methods for the TV-VAR(2) simulated data.}\label{table:32}
\begin{center}
 \begin{tabular}{c c c c}  
 \hline
 Model   & Case  1 & Case  2  & Case  3 \\  
 \hline
 TV-VPARCOR &  1.046(0.332) & 1.061(0.327) & 1.103(0.371) \\
 BCLF       &  1.038(0.317) & 1.046(0.325) & 1.085(0.354) \\
 \hline
\end{tabular}
\end{center}
\end{table}

\newpage

\subsection{Simulation 2: 20-Dimensional TV-VAR(1)}\label{sec:sim2}
We consider a simulation from \cite{zhao2019effcient} and generate data from a 20-dimensional nonstationary TV-VAR(1) process of length T = 300. The elements $\Phi_{i,j,t}$ of the autoregressive coefficients at time $t$, $\Phi_t$, are given as follows:
\[ \phi_{i,j} =
  \begin{cases} 
    0.7 + \frac{0.2}{299}\times t   & \text{for all} \quad   i=j,i=1,\dots,10,\\
    0.7 + \frac{0.2}{299}\times t   & \text{for all} \quad   i=j,i=1,\dots,10,\\
  \qquad  \quad 0.9                       & \text{for}  \; \qquad (i,j) \in [(1,5),(2,15)]\\
    \qquad -0.9                      & \text{for}  \; \qquad (i,j) \in [(6,12),(15,20)]\\
    \qquad \quad  0                       & \text{otherwise,}    
  \end{cases}
\]
for $t=1,\dots,300$. Additionally, the innovation covariance is specified as $\Sigma=0.1\bfI$.

\begin{table}
\caption{Computation times (in seconds) of fitting one simulated dataset for Simulation 1 and 2 using  TV-VPARCOR and BCLF models. The times cover the selection of discount factors, the selection of the model order, and the estimation of parameters. Note that the TV-VPARCOR method is conducted using Rcpp and parallel computing while the BCLF is coded in R. }\label{table:33}
\begin{center}
 \begin{tabular}{r r r} 
 \hline
 Model   & Simulation 1 & Simulation 2  \\  
 \hline
 TV-VPARCOR~(Rcpp) & 10.83 seconds &  17874.36 seconds\\
 BCLF~(R)       & 9.86 seconds& 279.76 seconds\\
 \hline
\end{tabular}
\end{center}
\end{table}

\clearpage

\LTcapwidth=\textwidth
\begin{longtable}{c c c c} 
\caption{Mean ASEs and corresponding standard deviations (in parentheses) using different methods for 500 simulated datasets from TV-VAR(2).}\label{table:34}\\

 \hline 
 Model  &  $g_{11}$  & $g_{22}$ & $\rho_{12}^2$ \\ [0.5ex] 
 \hline
 \multicolumn{4}{c}{Case  1: $\phi_{1,1,2,t}=0, \phi_{2,1,2,t}=0, \Sigma_t = I_2$} \\
 TV-VAR   &  0.05393~(0.02349) & 0.05497~(0.04214) & 0.00292~(0.00194) \\
 TV-VPARCOR &  0.03302~(0.01313) & 0.04062~(0.02288) & 0.00353~(0.00424) \\
 BCLF       &  0.04037~(0.01905) & 0.03643~(0.01713) & 0.00123~(0.00107) \\
  \hline
 \multicolumn{4}{c}{Case  1: $\phi_{1,1,2,t}=0, \phi_{2,1,2,t}=0, \Sigma_t = 2I_2$} \\
 TV-VAR   &  0.07538~(0.03663) & 0.07748~(0.0852) &  0.00484~(0.00357) \\
 TV-VPARCOR &   0.04144~(0.02542) & 0.04960~(0.0305) & 0.00582~(0.00595) \\
 BCLF       &   0.03509~(0.01680) & 0.03622~(0.0169) & 0.00124~(0.00112) \\
  \hline
 \multicolumn{4}{c}{Case  1: $\phi_{1,1,2,t}=0, \phi_{2,1,2,t}=0, \Sigma_t = 3I_2$} \\
  TV-VAR   &  0.10443~(0.05255) & 0.10575~(0.05543) &  0.00643~(0.00574) \\
 TV-VPARCOR &   0.05114~(0.04774)	 & 0.06347~(0.05972) & 0.00804~(0.00877) \\
 BCLF       &   0.03513~(0.01634) & 0.03623~(0.01699) & 0.00124~(0.00111) \\
 \hline
 \multicolumn{4}{c}{Case  2: $\phi_{1,1,2,t}=-0.8, \phi_{2,1,2,t}=0, \Sigma_t = I_2$} \\
 TV-VAR  &   0.09164~(0.08122) & 0.0870~(0.070740) & 0.00583~(0.00364) \\
 TV-VPARCOR &   0.06124~(0.02644) & 0.0475~(0.021017) & 0.00491~(0.00480) \\
 BCLF       &   0.03712~(0.01255) & 0.0384~(0.012639) & 0.00262~(0.00133) \\
  \hline
 \multicolumn{4}{c}{Case  2: $\phi_{1,1,2,t}=-0.8, \phi_{2,1,2,t}=0, \Sigma_t = 2I_2$} \\
 TV-VAR  &  0.10632~(0.07053) & 0.09847~(0.07628) & 0.00733~(0.00551) \\
 TV-VPARCOR &   0.19222~(0.30678) & 0.12383~(0.19629) &  0.01293~(0.01698)\\
 BCLF       &  0.03833~(0.01303) & 0.03937~(0.01285) & 0.00259~(0.00125)\\
  \hline
 \multicolumn{4}{c}{Case  2: $\phi_{1,1,2,t}=-0.8, \phi_{2,1,2,t}=0, \Sigma_t = 3I_2$} \\
 TV-VAR  & 0.26823~(0.96065)  & 0.16537~(0.28518) & 0.00978~(0.00817) \\
 TV-VPARCOR &0.28447~(0.50113)  &  0.17805~(0.33872) & 0.01791~(0.02220) \\
 BCLF       &0.03836~(0.01304)  & 0.03937~(0.01285)  &  0.00258~(0.00125) \\
 \hline
 \multicolumn{4}{c}{Case  3: $\phi_{1,1,2,t}=r_{3,t},\phi_{2,1,2,t}=r_{4,t}, \Sigma_t = I_2$} \\
TV-VAR  &   0.09722~(0.09334) & 0.09132~(0.07920) & 0.00493~(0.00310) \\
 TV-VPARCOR &   0.11324~(0.12927) & 0.07579~(0.07308) & 0.00807~(0.01014) \\
 BCLF       &   0.06804~(0.05361) & 0.03902~(0.01943) & 0.00442~(0.00372) \\
  \hline
 \multicolumn{4}{c}{Case  3: $\phi_{1,1,2,t}=r_{3,t},\phi_{2,1,2,t}=r_{4,t}, \Sigma_t = 2I_2$} \\
 TV-VAR  &  0.10583~(0.07045)  & 0.10023~(0.08221) & 0.00622~(0.00492) \\
 TV-VPARCOR &  0.19213~(0.30681) & 0.12378~(0.19631) &  0.01292~(0.01698)\\
 BCLF       &  0.06968~(0.05467) & 0.03996~(0.01970) &  0.00446~(0.00377)\\
  \hline
 \multicolumn{4}{c}{Case  3: $\phi_{1,1,2,t}=r_{3,t},\phi_{2,1,2,t}=r_{4,t}, \Sigma_t = 3I_2$} \\
 TV-VAR  & 0.19117~(0.68771) & 0.15403~(0.28426)& 0.00836~(0.00907) \\
 TV-VPARCOR &  0.28374~(0.49940) &  0.17709~(0.33651) &  0.01791~(0.02219)\\
 BCLF       &  0.06980~(0.05447) & 0.04005~(0.01978) & 0.00445~(0.00378)\\
  \hline

\end{longtable}

\begin{table}
\caption{Mean ASE values and corresponding standard deviations (in parentheses) for 500 simulated datasets from TV-VAR(2).}\label{table:35}
\begin{center}
 \begin{tabular}{c c c c} 
 \hline
 \multicolumn{4}{c}{Case 4: $\phi_{1,1,2,t}=0, \phi_{2,1,2,t}=0$ } \\
 Model  &  $g_{11}$  & $g_{22}$ & $\rho_{12}^2$ \\ [0.5ex] 
 TV-VAR  &  0.0916~(0.0812) & 0.0870~(0.0707) & 0.0058~(0.0036) \\
 TV-VPARCOR &  0.0614~(0.0264) & 0.0475~(0.0210) & 0.0049~(0.0048) \\
 BCLF       &   0.0371~(0.0126) & 0.0384~(0.0126) & 0.0026~(0.0013) \\
  \hline
  \multicolumn{4}{c}{Case 5: $\phi_{1,1,2,t}=-0.8, \phi_{2,1,2,t}=0$ } \\
 Model  &  $g_{11}$  & $g_{22}$ & $\rho_{12}^2$ \\ [0.5ex] 
 TV-VAR & 0.1174~(0.0819) & 0.1232~(0.0701) & 0.0056~(0.0034) \\
 TV-VPARCOR & 0.0938~(0.0296) & 0.0871~(0.0204) & 0.0048~(0.0045) \\
 BCLF       & 0.0396~(0.0129) & 0.0427~(0.0130) & 0.0027~(0.0012) \\
  \hline
\multicolumn{4}{c}{Case 6: $\phi_{1,1,2,t}=r_{3,t},\phi_{2,1,2,t}=r_{4,t}$} \\
 Model  &  $g_{11}$  & $g_{22}$ & $\rho_{12}^2$ \\ [0.5ex] 
 TV-VAR  &  0.1228~(0.0934) & 0.1281~(0.0782) & 0.0047~(0.0029) \\
 TV-VPARCOR &  0.1465~(0.1270) & 0.1159~(0.0736) & 0.0073~(0.0087) \\
 BCLF       &  0.0788~(0.0660) & 0.0449~(0.0206) & 0.0051~(0.0040) \\
  \hline
\end{tabular}
\end{center}
\end{table}

\begin{figure}
\caption{Top: True log spectral densities of Components 1, 2, 8, and 15. Bottom: estimated log spectral densities of Components 1, 2, 8, and 15 using the BCLF.}
\label{fig:327}
\begin{center}
\includegraphics[height=6in,width=6in]{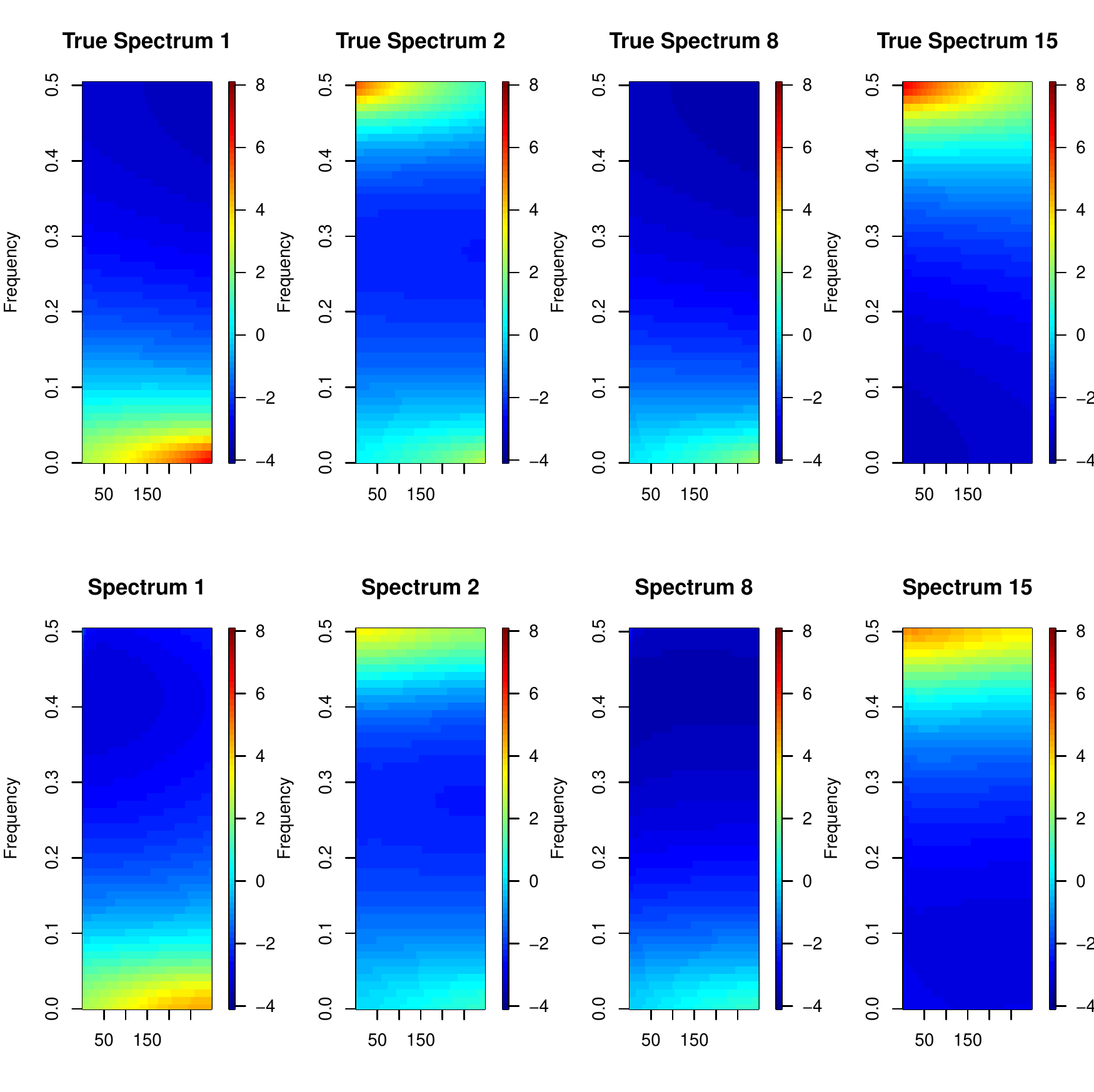}
\end{center}
\end{figure}

\begin{figure}
\caption{Top: True coherence between Components 1 and 5, 2 and 15, 5 and 12, and 15 and 20. Bottom: Estimated coherences between Components 1 and 5, 2 and 15, 5 and 12, and 15 and 20 using the BCLF.}
\label{fig:328}
\begin{center}
\includegraphics[height=6in,width=6in]{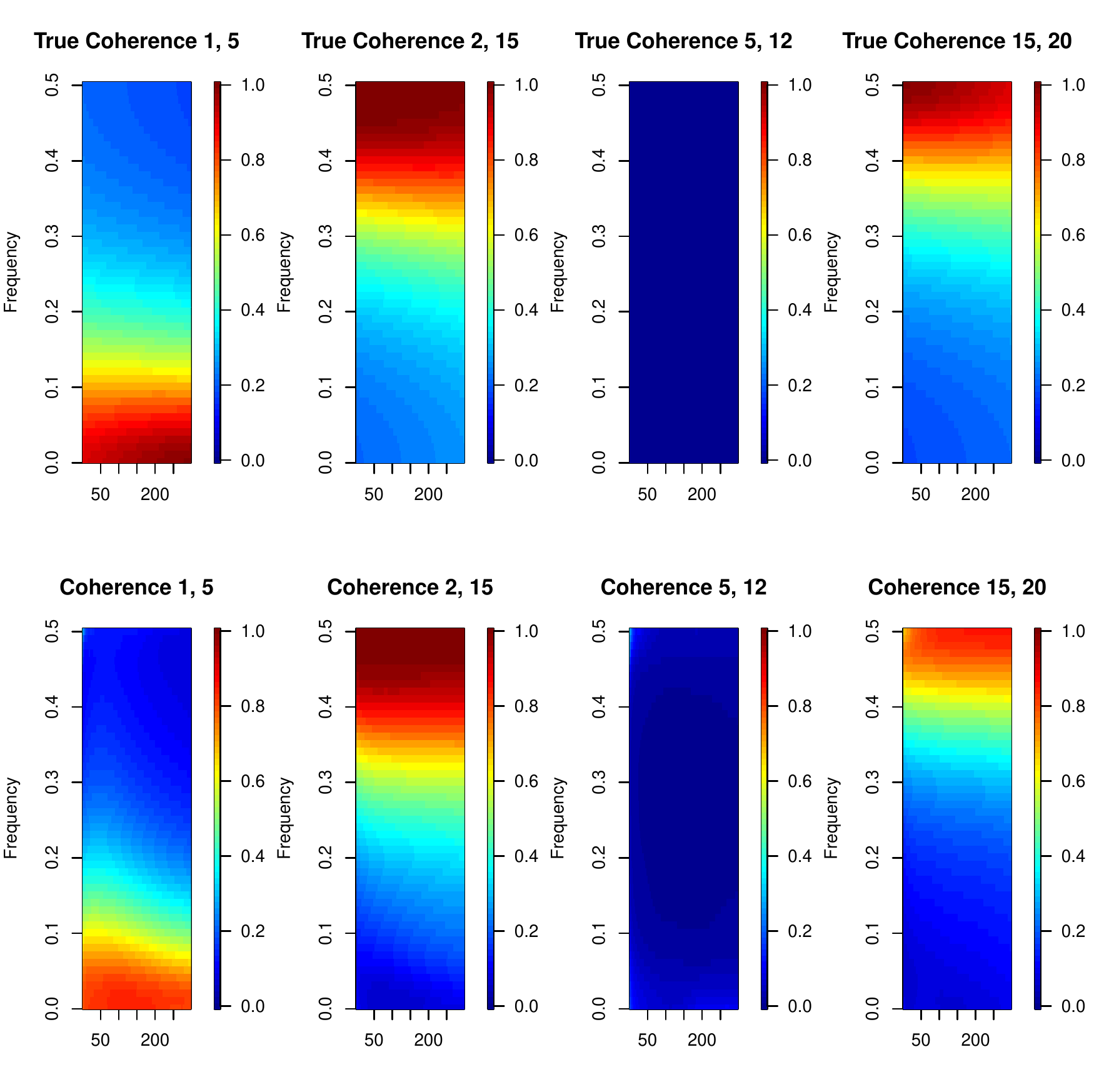}
\end{center}
\end{figure}

Similar to \cite{zhao2019effcient}, we choose a $P_{max}=3$ and fit the simulated datasets with the TV-VAR order adaptively selected based on BIC as we suggest. The other model selection criteria were also obtained. All of BIC, DIC, and WAIC suggest that TV-VAR(1) is the best model for the simulated datasets. The discount factors are chosen from a grid of values in $[0.99, 1]$ based on the likelihood of the fitted TV-AR models in each stage. Figures~\ref{fig:327} and \ref{fig:328} summarize the posterior inference obtained from the BCLF approach on the 100 simulated datasets. These figures show the average of the estimated spectrum of the two series and the average of the estimated coherence between them over the 100 simulated datasets. Table~\ref{table:33} shows the computation time of the simulation studies, including the selection of discount factors, the selection of the model order and the estimation of parameters. According to the results in this simulation studies, the BCLF shows superior performance relative to the other models considered in terms of both parameter estimation and forecasting. Importantly, at the same time, we see unmatched computational efficiency.

\section{Case Studies}\label{sec:application}

\subsection{Quarterly GDP Data}
We apply the BCLF to the quarterly change in seasonally adjusted GDP (in percentage) for five countries: the United States (US), Canada (CA), United Kingdom (UK), South Korea (KO), and Taiwan (TW) \citep{matteson2011dynamic}. We compare the out-of-sample forecast performance of our approach with standard vector autoregressive (VAR) models. The data can be obtained from the Organization for Economic Cooperation and Development (OECD) website (\url{https://www.oecd.org/}). The data we use begins with the first quarter of 1981 and goes through the second quarter of 2009. The time series plot of this dataset shows that the US, CA, and UK are more closely correlated than KO and TW. We want to use these correlated series to compare the prediction performance of the BCLF with standard VAR models. 

We set $P_{max} = 10$ and consider discount factor values on a grid in the $[0.90, 1]$ range and assume that each country has tits own optimal discount factor values. Throughout the rolling predictions, the BCLF always selected the order 1 as the best model order. Moreover, to address the problem of ordering uncertainty, we use the {\it Dynamic Ordering Selection} (DOS) and the {\it Dynamic Ordering Averaging} (DOA)  approaches proposed by \cite{levy2021dynamic} in conjunction with BCLF. We use the R package {\bf MTS} \citep{ruey2021MTS} to conduct VAR model prediction. All of AIC, BIC and the Hannan and Quinn information criterion suggest the best model order is 1. Table~\ref{table:36} shows the mean square prediction error (MSPE) of the one-step-ahead rolling predictions starting from the second quarter of 2004 through the second quarter of 2009.  From this table we can see that the BCLF provides superior prediction on this quarterly GDP data relative to the standard VAR model in terms of MSPE. In addition, the DOS and DOA reduce the ordering uncertainty and, thus, improves the forecast accuracy.

\begin{figure}
\caption{The quarterly percentage change in seasonally adjusted GDP of the United States (US), Canada (CA), United Kingdom (UK), Korea (KO), Taiwan (TW) from 1981QI through 2009QII. The vertical line is at 2004QII, which is the beginning of the out-of-sample one-step-ahead forecasts region.}
\label{fig:33}
\begin{center}
\includegraphics[height=5in,width=5in]{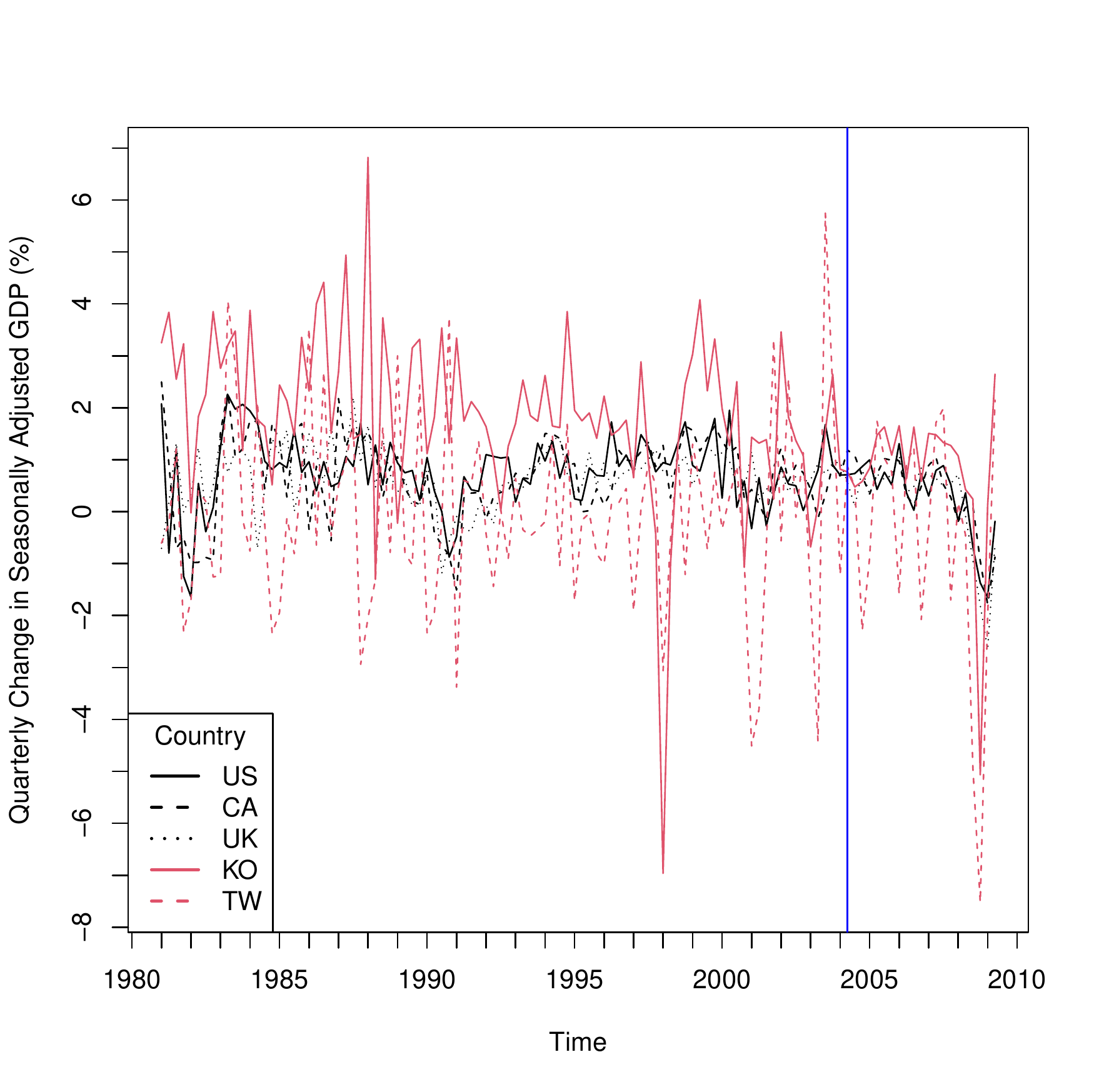}
\end{center}
\end{figure}

\begin{table}[h]
\caption{MSPE of rolling one-step-ahead prediction on the quarterly GDP Data. BCLF-DOS and BCLF-DOA denote the BCLF using DOS and DOA to solve ordering uncertainty, respectively. VAR denotes the standard VAR model. Both the VAR model with mean vector and without mean vector are included.}
\begin{center}\label{table:36}
 \begin{tabular}{ r| r } 
 \hline
 Method   & MSPE\\  
 \hline
BCLF      & 2.086 \\
BCLF-DOS  & 1.939 \\
BCLF-DOA  & 1.944 \\
VAR without mean vector & 2.249 \\
VAR with mean vector  & 2.232\\
 \hline
\end{tabular}
\end{center}
\end{table}

\subsection{Wind Data}
\begin{figure}
\caption{Top Row: Estimated log-spectral densities of the $X$ (East-West) components for Monterey (Spectrum 1), Salinas (Spectrum 2), and Watsonville (Spectrum 3). Bottom Row: Estimated log-spectral densities of the $Y$ (North-South) components for Monterey (Spectrum 4), Salinas (Spectrum 5), and Watsonville (Spectrum 6). The unit of time in the plots is 4 hours.}
\begin{center}
\includegraphics[height=4in,width=6.5in]{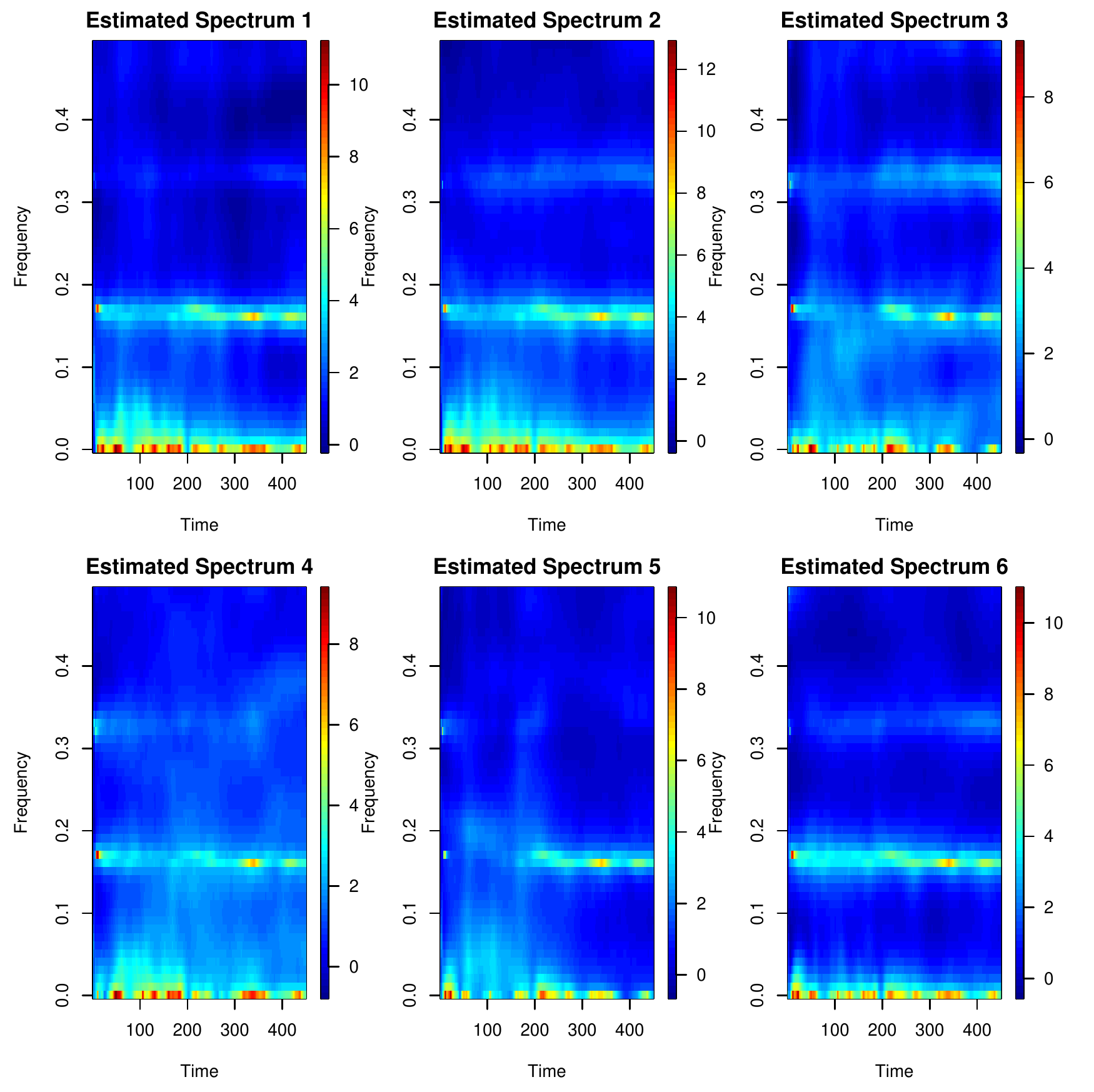}
\label{fig:39}
\end{center}
\end{figure}

\begin{figure}
\caption{Top Row: Estimated squared coherences between the $X$ (East-West) component and $Y$ (North-South) component in Monterey, Salinas, and Watsonville. Middle Row: Estimated squared coherences between the $X$ (East-West) components of Monterey and Salinas, Monterey and Watsonville, and Salinas and Watsonville. Bottom Row: Estimated squared coherences between the $Y$ (North-South) components of Monterey and Salinas, Monterey and Watsonville, and Salinas and Watsonville. The unit of time in the plots is 4 hours.}
\begin{center}
\includegraphics[height=6in,width=6.5in]{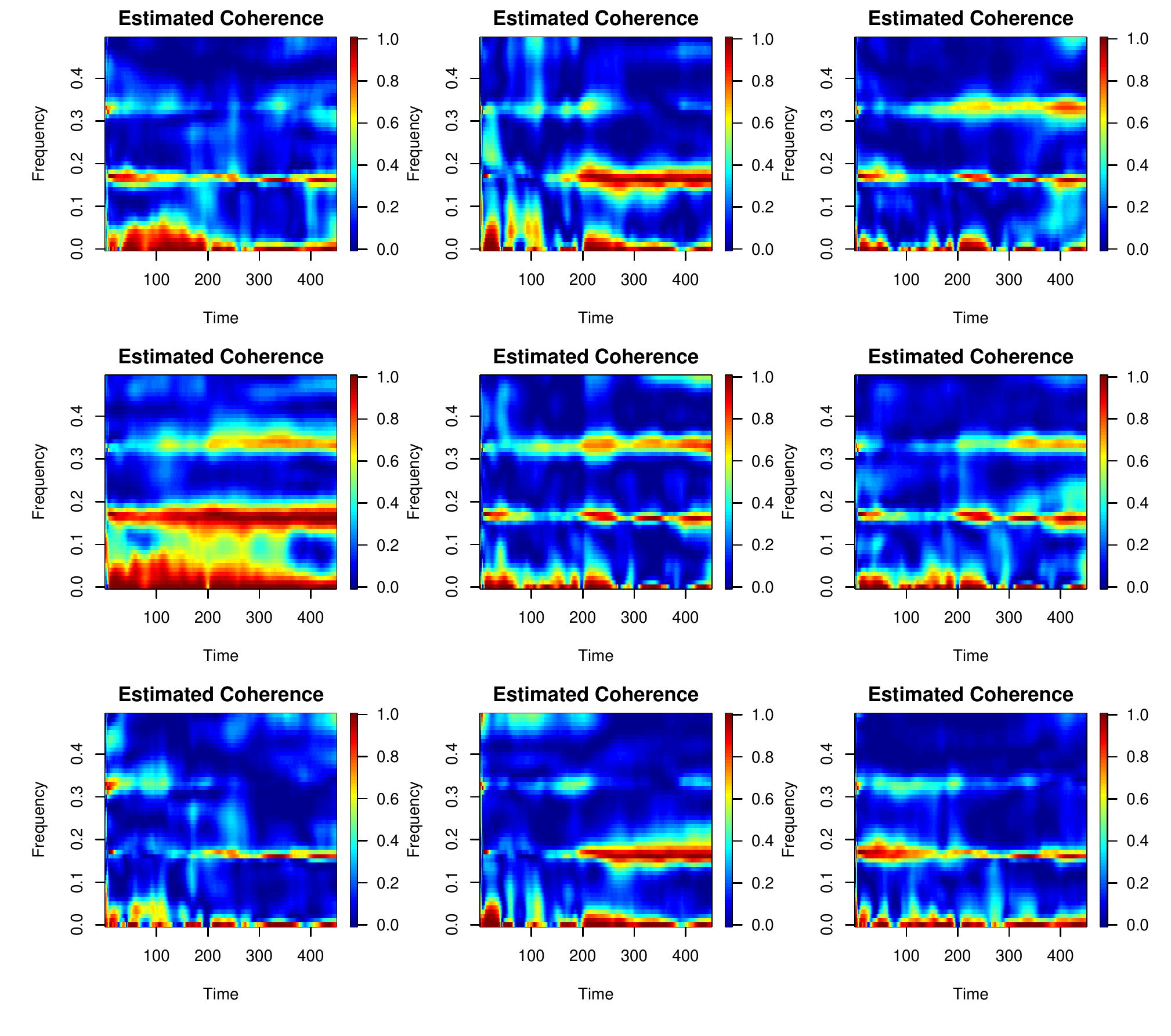}
\label{fig:310}
\end{center}
\end{figure}

We apply the BCLF to the wind data used by \cite{zhao2019effcient} and conduct a time-frequency analysis and make 72 hour forecasts. This dataset contains the median wind speed and direction measurements  collected every 4 hours  from 6/1/2010 to 8/15/2020 in Monterey, Salinas, and Watsonville, 3 stations located near Monterey Bay, Northern California. These data can be obtained from a publicly available database, the Iowa Environmental Mesonet (IEM) (see \url{http://mesonet.agron.iastate.edu/ASOS/}). ASOS stations are located at airports and take observations and basic reports from the National Weather Service (NWS), the Federal Aviation Administration (FAA), and the Department of Defense (DOD). We use the BCLF approach for a multivariate analysis of the six-dimensional time series of the wind speed component and wind direction component for the 3 stations. We set $P_{max} = 10$ and consider discount factor values on a grid in the $(0.95, 0.99]$ range and assume that each station has their own optimal discount factor values. BIC selects the model order 6 for the TV-VAR model. After applying BCLF to fit the TV-VAR(6) model to the six-dimensional data, we obtain the estimate of its time-frequency representation. Figure~\ref{fig:39} shows the estimated log spectral densities of the $X$ (East-West) component and the $Y$ (North-South) component  for each location. Figure~\ref{fig:310} shows the estimated squared coherence between each pair of wind components across the three locations. Generally, the results reveal a 24-hour cycle in $X$ components and $Y$ components and their coherence and the magnitude of the cycles vary with time.

The TV-VPARCOR model can also be used for forecasting as described in Section~\ref{sec:forecast}. Table~\ref{table:37} compares the 72 hour forecasts obtained from the Naive, TV-VAR, TV-VPARCOR and BCLF model for the north–south wind component and east-west component in all three locations. Note that Naive denotes the naive forecast, that is, using the previous period to forecast for the next period (carry-forward). We see that the BCLF gives best forecasts among the four models in terms of the mean squared prediction error (MSPE).

Moreover, the computation time of each model varies significantly for this wind data. As we discussed previously, the BCLF is the fastest among those considered herein, the TV-VPARCOR model is second fastest, and the TV-VAR is significantly slower. 

\begin{table}[h]
\caption{MSPE of rolling one-step-ahead prediction on the Wind Data for next 72-hours}
\begin{center}\label{table:37}
 \begin{tabular}{r| r } 
 \hline
 Method   & MSPE \\  
 \hline
BCLF      & 7.2163  \\
TV-PARCOR & 7.8931 \\
TV-VAR    & 9.4009 \\
Naive     & 14.2487 \\
 \hline
\end{tabular}
\end{center}
\end{table}

\section{Discussion}\label{sec:discussion}
We propose a computationally efficient model-based parametric approach for nonstationary multivariate time series. We use a response-orthogonal reparameterization to transform a TV-VAR model into a scalar periodic AR model and further use the Bayesian Lattice filter together with PARCOR to speed up the estimation. Our approach can simultaneously estimate the time-varying coefficients and the time-varying innovation covariance. The ``one channel at-a-time" modeling avoids matrix inversion computations in the estimation, which significantly reduces the computation cost and makes the computation time increase linearly, rather than exponentially, with the data dimension. Additionally, this ``one channel at-a-time" modeling makes it possible for our approach to benefit from parallel computing, which is a subject of future research. Moreover, the use of Bayesian lattice filters makes the computation time increase linearly, rather than exponentially, with the model order. We also provide a model selection method to choose the discount factors and the optimal model order. The simulation cases show that the parameter estimation is fairly accurate. The GDP and Wind data examples shows that the time-varying coefficients and innovation covariance can be effectively used to reveal the time-varying structure.  

Our method benefits from the LDL decomposition (modified Cholesky decomposition) of the innovation covariance as in \eqref{eq:A2}. This framework introduces a problem of order uncertainty \citep{primiceri2005time, zhao2016dynamic, lopes2021parsimony}. In particular, the ordering of the multiple variables matters because any permutation of the ordering can lead to different prior distributions for the AR coefficients. The prior distributions are firstly assigned to the PARCOR coefficients after making a transformation through the Cholesky decomposition. When transforming back to the original model space, these priors are transformed to the priors of the AR coefficients. This transformation depends on the order of the variables in the multivariate model. This difference in the prior distributions does not have a significant effect on the time-invariant parameter models. However, it may show a more noticeable difference in the time-varying parameter models when using different ordering. In Simulation~1, we see this phenomenon. However, in both our motivating data applications and in \cite{primiceri2005time}, different orders of the variables led to very similar results. \cite{levy2021dynamic} provide a comprehensive discussion on the ordering uncertainty and proposes two solutions: DOS and DOA. We adopt both in the quarterly GDP data application to address the ordering uncertainty and find that it improves the forecast accuracy.

\section*{Acknowledgements}

 This research was partially supported by the U.S.~National Science Foundation (NSF) under NSF grant SES-1853096. This article is released to inform interested parties of ongoing research and to encourage discussion. The views expressed on statistical issues are those of the authors and not those of the NSF or U.S. Census Bureau. The authors thank Drs. Matteson and Zhao for their assistance with the quarterly GDP data and California Wind data, respectively. \vspace*{-8pt}


\appendix

\section{Algorithm for Fitting Multivariate Time Series}\label{sec:appendixa}
We summarize the algorithm of our approach to  fitting a TV-VAR model as follows:
\begin{itemize}
\item{Step 1}. Interlace the multivariate times series as in \eqref{eq:1} into a periodic time series as in \eqref{eq:3}.

\item{Step 2}. Set up a value for order $P$ and a set of values for $\gamma_{k,m}$, $\delta_{k,m}$ for $m = 1,\dots, M_k$, $k=1,\dots,K$, where $M_k = kP_{max}+K-1$, as well as the initial values of parameters at $t$ = 0. 

\item{Step 3}. Repeat Step 4 for stage $m=1,\dots,M_k$, $k=1,\dots,K$.

\item{Step 4}. Apply the sequential filtering and smoothing algorithm to the prediction errors of last stage, $f^{(m-1)}_{k,t}$ and $b^{(m-1)}_{k,t}$ to obtain $\alpha_{t}^{(m)}=\mu_{t,T}^{(m)}$ and $\sigma_{t}^{2(m)}=s_{t,T}^{(m)}$ of the forward and backward equations, and the forward and backward prediction errors, $f^{(m)}_{k,t}$ and $b^{(m)}_{k,t}$ for $t=1,\dots,T$.

\item{Step 5}. To compute the model selection criterion of TV-VAR($P$), collect the computed parameters up to order $kP+K-1$ for the $k$th series, $k=1,2,\dots,K$. Transform the computed parameters to obtain the estimated values of the parameters $\{\bfPhi_{p,t}\}$ and $\{\bfSigma_t\}$ in \eqref{eq:1}. Compute the criterion following Section~\ref{ms}. Pick the best order according to the model selection criterion.

\item{Step 6}. (Optional) Compute the spectrum and cross-spectrum based on the parameter estimation following Section~\ref{sec:sim1}.

\item{Step 7}. (Optional) Make $h$-step ahead forecasting following Section~\ref{sec:forecast}.

\end{itemize}

\section{Sequential Filtering and Smoothing Algorithm}\label{sec:appendixb}
Suppose we need to fit \eqref{eq:fb1} and \eqref{eq:fb2} to estimate the PARCOR coefficients and innovation variance. Their posterior distributions can be obtained by following a sequential filtering and smoothing algorithm. The algorithm given here is for the forward autoregression case as in \eqref{eq:fb1}. Filtering and smoothing algorithm can be obtained for the backward case in a similar manner. For any series, any stage, we denote the posterior distribution at time $t$ as $(\alpha_t| D_t) \sim T_{\nu_t}(\mu_t,C_t)$ a multivariate T distribution with $\nu_t$ df, location parameter $\mu_t$, and scale matrix $C_t$, and $(\sigma_t^{-2}| D_t) \sim G(\nu_t/2,\kappa_t/2)$, a gamma distribution with shape parameter $\nu_t/2$ and scale parameter $\kappa_t/2$. These parameters can be computed for all $t$ using the filtering equations below. Note we use $s_t=\kappa_t/\nu_t$ to denote the usual point estimate of $\sigma_t^2$. $f_t$ in the equation is the forward prediction error. For $t=2,\dots,T$, we have
\begin{align*}
    \mu_t &= \mu_{t-1} + z_t e_t,\\
    C_t &= (R_t-z_t z_t'q_t)(s_t/s_{t-1}),
\end{align*}
and
\begin{align*}
    \nu_t &= \delta \nu_{t-1} + 1,\\
    \kappa_t &= \delta \kappa_{t-1} + s_{t-1}e_t^2/q_t ,
\end{align*}
where
\begin{align*}
    e_t &= f_t- z_{t-1}' m_{t-1}, \\
    q_t &= z_{t-1}'R_t z_{t-1}'+ s_{t-1},
\end{align*}
and
\begin{align*}
    z_t &= R_t f_{t-1}/q_t, \\
    R_t &= C_{t-1} + G_t, \\
    G_t &= C_t (1-\beta)/\beta.
\end{align*}
After the filtering equations up to $T$, we compute the full marginal posterior distribution $(\alpha_t| D_T) \sim T_{\nu_t}(\mu_{t,T},C_t)$ and $(\sigma_t^{-2}| D_T) \sim G(\nu_{t,T}/2,\kappa_{t,T}/2)$ through the smoothing equations
\begin{align*}
    \mu_{t,T} &= (1-\beta)\mu_t + \beta \mu_{t+1,T}\\
    C_{t,T} &= [(1-\beta)C_t + \beta^2 C_{t+1,T}](s_{t,T}/s_t)\\
    \nu_{t,T} &= (1-\delta)\nu_t + \delta \nu_{t+1,T}\\
    1/s_{t,T} &= (1-\delta)/s_t + \delta s_{t+1,T}
\end{align*}
and $\kappa_{t,T}=\nu_{t,T}s_{t,T}$ for $t=T-1,\dots,1$.

\section{Model Selection Criteria}\label{sec:appendixc}
{\bf Bayesian Information Criterion (BIC):} For a TV-VAR$(P)$ model with parameters denoted as $\bftheta$, BIC is defined as  
$$ BIC(P) = -2 \Lagr + n_\Theta log(KT). $$
BIC is computed for each of the fitted models of order $P=1,\dots,P_{max}$ to a $K$-dimensional time series of length $T$. $\Lagr$ is the log likelihood based on multivariate time-varying AR model of order $P$ as \eqref{eq:1}. $n_\Theta $ is the number of parameters estimated in the initial state of the state space model. In our case, $n_\Theta = 2PK^2+(K-1)K$, which is the total number of estimated time-varying variables in all stages. We use BIC for model selection criterion throughout this paper. BIC works well for identifying the correct order on simulated examples.\\~\\
{\bf Deviance Information Criterion (DIC):} With parameters denoted as $\bftheta$ and data denoted as $\bfy$, the DIC is defined as
\begin{equation*}
    DIC = -2 \text{log}p(\bfy|\widehat{\bftheta}_{Bayes})+2p_{DIC},
\end{equation*}
where $\widehat{\bftheta}_{Bayes}$ is the Bayes estimator of $\bftheta$ and $p_{DIC}$ is the effective number of parameters. The effective number of parameters is given by
\begin{equation*}
    p_{DIC} = 2 \left[ \text{log}p(\bfy|\widehat{\bftheta}_{Bayes})-E_{post}(\text{log}p(\bfy|\bftheta))\right],
\end{equation*}
where $E_{post}(\cdot)$ is the expectation under the posterior distribution. In our case, the terms above are estimated through MC samples $\bftheta^{(s)}$ as
\begin{equation*}
    p_{DIC} = 2 \left[ \text{log}p(\bfy|\widehat{\bftheta}_{Bayes})-\frac{1}{S}\overset{S}{\underset{s=1}{\sum}}(\text{log}p(\bfy|\bftheta^{(s)}))\right],
\end{equation*}
where samples $\bftheta^{(s)}$, $s = 1, \dots, S$, are generated from $p(\bftheta|\bfy)$, the posterior distribution of $\bftheta$ by the following steps: 
\begin{enumerate}
\item  generate samples $\{\alpha_t\}$ from $T_{\nu_t}(\mu_t,C_t)$ and samples $\{\sigma_t^{-2}\}$ from $G(\nu_t/2,\kappa_t/2)$ for $t=1,\dots,T$ as Section~\ref{sec:appendixb};

\item  transform them into TV-VAR parameters, $\{\bfPhi_{p,t}\}$ and $\{\bfSigma_t\}$, by \eqref{eq:A1} and \eqref{eq:A2};

\item  generate $S$ sets of samples $\{\bftheta^{(s)} = (\bfPhi_{t,p}^{(s)}, \bfSigma_t^{(s)}), s=1,\dots,S\}$.               
\end{enumerate}

\noindent {\bf Widely Applicable Akaike Information Criterion (WAIC):} WAIC is a generalized version of AIC and is more appropriate for Bayesian hierarchical models \citep{watanabe2010asymptotic}. For a model with parameters denoted as $\bftheta$, considering data $\bfy=(y_1,\dots,y_T)'$, the WAIC is defined as
\begin{equation*}
    WAIC = -2 \text{log}p(\bfy|\widehat{\bftheta}_{Bayes})+2p_{WAIC},
\end{equation*}
where
\begin{equation*}
    p_{WAIC} = 2 \overset{T}{\underset{t=1}{\sum}} \left( \text{log}(E_{post}p(y_t|\bftheta)-E_{post}(\text{log}p(y_t|\bftheta))\right).
\end{equation*}
WAIC is estimated by using MC samples from the S
posterior draws $\bftheta^{(s)}$ as
\begin{equation*}
    \widehat{p}_{WAIC} = 2 \overset{T}{\underset{t=1}{\sum}} \left( \text{log}(\frac{1}{S}\overset{S}{\underset{s=1}{\sum}}p(y_t|\bftheta^{(s)})-\frac{1}{S}\overset{S}{\underset{s=1}{\sum}}(\text{log}p(y_t|\bftheta^{(s)}))\right).
\end{equation*}
$\bftheta^{(s)}$ is generated from $p(\bftheta|\bfy)$, the posterior distribution of $\bftheta$ in the same way as DIC.

To achieve a best model selection, we first search for an order which achieves the smallest BIC, DIC, WAIC. If the numerical criteria, BIC, DIC, WAIC, do not agree with each other, we use the visual scree plot to assist the model selection. In practice, for BCLF, BIC works best for selecting the model order among the numerical criteria in most of the simulation cases we conducted.

\section{Forecasting}\label{sec:appendixd}
We can make $h$-step ahead forecasting by following these steps.
\begin{itemize}
    \item For stage $m=1,\dots,M_k$ and series $k=1,\dots,K$, compute the $h$-step ahead posterior predictive distribution of the PARCOR coefficients following \cite{west1997bayesian}:  $(\alpha_{f,m,k+(T+h-1)K}^{(m)}|D_T) \sim N(\mu_{f,m,k+(T-1)K}^{(m)}(h),C_{f,m,k+(T-1)K}^{(m)}(h))$ where $ \mu_{f,m,k+(T-1)K}(h) = \mu_{f,m,k+(T+h-1)K}^{(m)} $ and $ C_{f,m,k+(T-1)K}^{(m)}(h) = C_{f,m,k+(T-1)K}^{(m)} + h G_{f,m,k+TK}^{(m)}$ with $G_{f,m,k+TK}^{(m)} = C_{k+(T+h-1)K}^{(m)} (1-\beta)/\beta$. 
    \item For stage $m=1,\dots,M_k$ and series $k=1,\dots,K$, draw $J$ samples for each of the $h$-step ahead of the AR coefficients $\{a_{m,k+(T+h-1)K}^{(M_k)}$, $m=1,\dots,M_k\}$ from the samples of $\{\alpha_{f,m,k+(T+h-1)K}^{(m)}\}$ and $\{\alpha_{b,m,k+(T+h-1)K}^{(m)}\}$ for series $k=1,\dots,K$. 
    \item Transform the samples of $\{a_{m,k+(T+h-1)K}^{(M_k)}$, $m=1,\dots,M_k\}$ into the samples of TV-VAR parameters $\{\bfPhi_{p,T+h},p=1,\dots,P\}$.
    \item The samples of the $h$-step ahead forecast is obtained as 
       \begin{equation}\label{eq:ch3f2}
          \bfx_{T+h}^{(j)} =  \overset{P}{\underset{p=1}{\sum}}   \bfPhi_{p,T+h}^{(j)} \bfx_{T+h-p}^{(j)}, 
        \end{equation}
       where $\bfx_{T+h-p}^{(j)} = \bfx_{T+h-p}$ if $h-p\le 0$.
       
    \item We use the posterior mean of $\bfx_{T+h}$ obtained through the samples in \eqref{eq:ch3f2} as the $h$-step ahead forecast.
\end{itemize}

\clearpage

\bibliography{biblio}

\begin{thebibliography}{35}
\newcommand{\enquote}[1]{``#1''}
\expandafter\ifx\csname natexlab\endcsname\relax\def\natexlab#1{#1}\fi
\expandafter\ifx\csname url\endcsname\relax
  \def\url#1{{\tt #1}}\fi
\expandafter\ifx\csname urlprefix\endcsname\relax\def\urlprefix{URL }\fi

\bibitem[{Brockwell and Davis(2009)}]{brockwell1991time}
Brockwell, P.~J. and Davis, R.~A.
\newblock {\em Time Series: Theory and Methods\/}.
\newblock Springer, second edition (2009).

\bibitem[{Del~Negro and Primiceri(2015)}]{del2015time}
Del~Negro, M. and Primiceri, G.~E.
\newblock \enquote{Time varying structural vector autoregressions and monetary
  policy: a corrigendum.}
\newblock {\em The Review of Economic Studies\/}, 82(4):1342--1345 (2015).

\bibitem[{Fan and Yao(2008)}]{fan2008nonlinear}
Fan, J. and Yao, Q.
\newblock {\em Nonlinear Time Series: Nonparametric and Parametric Methods\/}.
\newblock Springer Science \& Business Media (2008).

\bibitem[{Gelman et~al.(2013)Gelman, Carlin, Stern, Dunson, Vehtari, and
  Rubin}]{gelman2013bayesian}
Gelman, A., Carlin, J.~B., Stern, H.~S., Dunson, D.~B., Vehtari, A., and Rubin,
  D.~B.
\newblock {\em Bayesian Data Analysis\/}.
\newblock Chapman \& Hall/CRC (2013).

\bibitem[{Gersch and Stone(1994)}]{gersch1994one}
Gersch, W. and Stone, D.
\newblock \enquote{One channel at-a-time multichannel autoregressive modeling
  of stationary and nonstationary time series.}
\newblock In Bozdogan, H., Sclove, S.~L., Gupta, A.~K., Haughton, D., Kitagawa,
  G., Ozaki, T., and Tanabe, K. (eds.), {\em Proceedings of the First US/Japan
  Conference on the Frontiers of Statistical Modeling: An Informational
  Approach: Volume 3 Engineering and Scientific Applications\/}, 165--192.
  Springer (1994).

\bibitem[{Gersch and Stone(1995)}]{gersch1995multivariate}
---.
\newblock \enquote{Multivariate autoregressive time semes modeling: one scalar
  autoregressive model at-a-time.}
\newblock {\em Communications in Statistics-Theory and Methods\/},
  24(11):2715--2733 (1995).

\bibitem[{Guo and Dai(2006)}]{guo2006multivariate}
Guo, W. and Dai, M.
\newblock \enquote{Multivariate time-dependent spectral analysis using Cholesky
  decomposition.}
\newblock {\em Statistica Sinica\/}, 16(3):825 (2006).

\bibitem[{Hayes(1996)}]{hayes1996sdspam}
Hayes, M.~H.
\newblock {\em Statistical Digital Signal Processing and Modeling\/}.
\newblock John Wiley \& Sons (1996).

\bibitem[{Huerta and Lopes(2000)}]{huerta2000bayesian}
Huerta, G. and Lopes, H.~F.
\newblock \enquote{Bayesian forecasting and inference in latent structure for
  the brazilian industrial production index.}
\newblock {\em Brazilian Review of Econometrics\/}, 20(1):1--26 (2000).

\bibitem[{Hunter et~al.(2017)Hunter, Burke, and
  Canepa}]{hunter2017multivariate}
Hunter, J., Burke, S.~P., and Canepa, A.
\newblock {\em Multivariate Modelling of Non-Stationary Economic Time
  Series\/}.
\newblock Springer (2017).

\bibitem[{Kitagawa(2010)}]{kitagawa2010timeseries}
Kitagawa, G.
\newblock {\em Introduction to Time Series Modeling\/}.
\newblock Chapman \& Hall/CRC (2010).

\bibitem[{Kitagawa and Gersch(1996)}]{kitagawa1996smoothness}
Kitagawa, G. and Gersch, W.
\newblock {\em Smoothness Priors Analysis of Time Series\/}.
\newblock Springer (1996).

\bibitem[{Kowal et~al.(2017)Kowal, Matteson, and Ruppert}]{kowal2017bayesian}
Kowal, D.~R., Matteson, D.~S., and Ruppert, D.
\newblock \enquote{A Bayesian multivariate functional dynamic linear model.}
\newblock {\em Journal of the American Statistical Association\/},
  112(518):733--744 (2017).

\bibitem[{Levy and Lopes(2021)}]{levy2021dynamic}
Levy, B.~P. and Lopes, H.~F.
\newblock \enquote{Dynamic Ordering Learning in Multivariate Forecasting.}
\newblock {\em arXiv preprint arXiv:2101.04164\/} (2021).

\bibitem[{Lopes et~al.(2021)Lopes, McCulloch, and Tsay}]{lopes2021parsimony}
Lopes, H.~F., McCulloch, R.~E., and Tsay, R.~S.
\newblock \enquote{Parsimony inducing priors for large scale state--space
  models.}
\newblock {\em Journal of Econometrics\/} (2021).

\bibitem[{Masry(1996)}]{masry1996multivariate}
Masry, E.
\newblock \enquote{Multivariate local polynomial regression for time series:
  uniform strong consistency and rates.}
\newblock {\em Journal of Time Series Analysis\/}, 17(6):571--599 (1996).

\bibitem[{Matteson and Tsay(2011)}]{matteson2011dynamic}
Matteson, D.~S. and Tsay, R.~S.
\newblock \enquote{Dynamic orthogonal components for multivariate time series.}
\newblock {\em Journal of the American Statistical Association\/},
  106(496):1450--1463 (2011).

\bibitem[{Nakajima et~al.(2011)Nakajima, Kasuya, and
  Watanabe}]{nakajima2011bayesian}
Nakajima, J., Kasuya, M., and Watanabe, T.
\newblock \enquote{Bayesian analysis of time-varying parameter vector
  autoregressive model for the Japanese economy and monetary policy.}
\newblock {\em Journal of the Japanese and International Economies\/},
  25(3):225--245 (2011).

\bibitem[{Nakajima and West(2013)}]{nakajima2013bayesian}
Nakajima, J. and West, M.
\newblock \enquote{Bayesian analysis of latent threshold dynamic models.}
\newblock {\em Journal of Business \& Economic Statistics\/}, 31(2):151--164
  (2013).

\bibitem[{Ombao et~al.(2001)Ombao, Raz, Von~Sachs, and Malow}]{ombao2001auto}
Ombao, H., Raz, J., Von~Sachs, R., and Malow, B.
\newblock \enquote{Automatic statistical analysis of bivariate nonstationary
  time series.}
\newblock {\em Journal of the American Statistical Association\/},
  96(454):543--560 (2001).

\bibitem[{Ombao et~al.(2005)Ombao, Von~Sachs, and Guo}]{ombao2005slex}
Ombao, H., Von~Sachs, R., and Guo, W.
\newblock \enquote{SLEX analysis of multivariate nonstationary time series.}
\newblock {\em Journal of the American Statistical Association\/},
  100(470):519--531 (2005).

\bibitem[{Pagano(1978)}]{pagano1978periodic}
Pagano, M.
\newblock \enquote{On periodic and multiple autoregressions.}
\newblock {\em The Annals of Statistics\/}, 6(6):1310--1317 (1978).

\bibitem[{Primiceri(2005)}]{primiceri2005time}
Primiceri, G.~E.
\newblock \enquote{Time varying structural vector autoregressions and monetary
  policy.}
\newblock {\em The Review of Economic Studies\/}, 72(3):821--852 (2005).

\bibitem[{Sakai(1982)}]{sakai1982circular}
Sakai, H.
\newblock \enquote{Circular lattice filtering using Pagano's method.}
\newblock {\em IEEE Transactions on Acoustics, Speech, and Signal
  Processing\/}, 30(2):279--287 (1982).

\bibitem[{Shephard(2005)}]{shephard2005stochastic}
Shephard, N.
\newblock {\em Stochastic Volatility: Selected Readings\/}.
\newblock Oxford University Press on Demand (2005).

\bibitem[{Shumway and Stoffer(2006)}]{shumway2006time}
Shumway, R.~H. and Stoffer, D.~S.
\newblock {\em Time Series Analysis and Its Applications (2nd ed)\/}.
\newblock Springer (2006).

\bibitem[{Sui(2021)}]{sui2021nonstationary}
Sui, Y.
\newblock \enquote{Nonstationary Bayesian Time Series Models with Time-Varying
  Parameters and Regime-Switching.}
\newblock Ph.D. thesis, University of Missouri (2021).

\bibitem[{Triantafyllopoulos(2007)}]{triantafyllopoulos2007covariance}
Triantafyllopoulos, K.
\newblock \enquote{Covariance estimation for multivariate conditionally
  Gaussian dynamic linear models.}
\newblock {\em Journal of Forecasting\/}, 26(8):551--569 (2007).

\bibitem[{Tsay and Wood(2021)}]{ruey2021MTS}
Tsay, R.~S. and Wood, D.
\newblock {\em MTS: All-Purpose Toolkit for Analyzing Multivariate Time Series
  (MTS) and Estimating Multivariate Volatility Models\/} (2021).
\newblock R package version 1.0.3.
\newline\urlprefix\url{https://CRAN.R-project.org/package=MTS}

\bibitem[{Watanabe(2010)}]{watanabe2010asymptotic}
Watanabe, S.
\newblock \enquote{Asymptotic equivalence of Bayes cross validation and widely
  applicable information criterion in singular learning theory.}
\newblock {\em Journal of Machine Learning Research\/}, 11(Dec):3571--3594
  (2010).

\bibitem[{West and Harrison(1997)}]{west1997bayesian}
West, M. and Harrison, J.
\newblock {\em Bayesian Forecasting and Dynamic Models (2nd)\/}.
\newblock Springer (1997).

\bibitem[{Yang et~al.(2016)Yang, Holan, and Wikle}]{yang2016bayesian}
Yang, W.-H., Holan, S.~H., and Wikle, C.~K.
\newblock \enquote{Bayesian lattice filters for time--varying autoregression
  and time--frequency analysis.}
\newblock {\em Bayesian Analysis\/}, 11(4):977--1003 (2016).

\bibitem[{Zhao(2022)}]{zhao2022efficient}
Zhao, W.
\newblock \enquote{Efficient Analysis of Multiple and Multivariate
  Non-stationary Time Series in the Partial Autocorrelation Domain.}
\newblock Ph.D. thesis, UC Santa Cruz (2022).

\bibitem[{Zhao and Prado(2020)}]{zhao2019effcient}
Zhao, W. and Prado, R.
\newblock \enquote{Effcient {Bayesian} PARCOR approaches for dynamic modeling
  of multivariate time series.}
\newblock {\em Journal of Time Series Analysis\/} (2020).

\bibitem[{Zhao et~al.(2016)Zhao, Xie, and West}]{zhao2016dynamic}
Zhao, Z.~Y., Xie, M., and West, M.
\newblock \enquote{Dynamic dependence networks: Financial time series
  forecasting and portfolio decisions.}
\newblock {\em Applied Stochastic Models in Business and Industry\/},
  32(3):311--332 (2016).

\end{thebibliography}
\bibliographystyle{jasa}

\end{document}